\PassOptionsToPackage{table}{xcolor}
\RequirePackage{pgfplots}
\usetikzlibrary{patterns} 

\documentclass[pdflatex, sn-apa]{sn-jnl}

\usepackage{cmap}						
\usepackage[T1]{fontenc}				
\usepackage[nameinlink]{cleveref}
\usepackage{csquotes}  
\usepackage{pdflscape} 

\usepackage{enumitem}
\usepackage{listings}
\usepackage{url} 

\newcommand{\numtotalpapers}{3021}
\newcommand{\numselectedpapers}{35}
\newcommand{\datesearchended}{September 24, 2021}
\newcommand{\datecitationanalysis}{October 8, 2021}

\renewcommand*{\arraystretch}{1.2}

\jyear{2022}
\newcommand\copyrighttext{%
  \footnotesize This version of the article has been accepted for publication, after peer review, but is not the Version of Record and does not reflect post-acceptance improvements or any corrections. The Version of Record is available online at: \url{https://doi.org/10.1007/s10639-022-11093-6}. Use of this Accepted Version is subject to the publisher’s Accepted Manuscript terms of use: \url{https://www.springernature.com/gp/open-research/policies/accepted-manuscript-terms}
  \smallskip
  
  \textcopyright\ 2022. Please cite this article as follows: V. Švábenský, J. Vykopal, P. Čeleda, L. Kraus: \textit{Applications of Educational Data Mining and Learning Analytics on Data From Cybersecurity Training}. Springer Education and Information Technologies, 2022, ISSN 1360-2357, DOI: \url{https://doi.org/10.1007/s10639-022-11093-6}}
\newcommand\copyrightnotice{%
\begin{tikzpicture}[remember picture,overlay]
\node[anchor=north,yshift=-24pt] at (current page.north) {\fbox{\parbox{\dimexpr\textwidth-\fboxsep-\fboxrule\relax}{\copyrighttext}}};
\end{tikzpicture}%
}

\begin{document}

\title[\tiny Applications of Educational Data Mining and Learning Analytics on Data From Cybersecurity Training]{Applications of Educational Data Mining and Learning Analytics on Data From Cybersecurity Training

\copyrightnotice\vspace*{-0.75cm}}

\author*[1,2]{\fnm{Valdemar} \sur{Švábenský}}\email{svabensky@ics.muni.cz}
\author[1]{\fnm{Jan} \sur{Vykopal}}\email{vykopal@ics.muni.cz}
\author[1]{\fnm{Pavel} \sur{Čeleda}}\email{celeda@ics.muni.cz}
\author[1]{\fnm{Lydia} \sur{Kraus}}\email{kraus@ics.muni.cz}

\affil*[1]{\orgdiv{Institute of Computer Science}, \orgname{Masaryk University}, \orgaddress{\street{Šumavská~15}, \city{Brno}, \postcode{60200}, \country{Czech Republic}}}
\affil*[2]{\orgdiv{Faculty of Informatics}, \orgname{Masaryk University},\\\orgaddress{\street{Botanická 68a}, \city{Brno}, \postcode{60200}, \country{Czech Republic}}}

\abstract{Cybersecurity professionals need hands-on training to prepare for managing the current advanced cyber threats. To practice cybersecurity skills, training participants use numerous software tools in computer-supported interactive learning environments to perform offensive or defensive actions. The interaction involves typing commands, communicating over the network, and engaging with the training environment. The training artifacts (data resulting from this interaction) can be highly beneficial in educational research. For example, in cybersecurity education, they provide insights into the trainees' learning processes and support effective learning interventions. However, this research area is not yet well-understood. Therefore, this paper surveys publications that enhance cybersecurity education by leveraging trainee-generated data from interactive learning environments. We identified and examined \numtotalpapers\ papers, ultimately selecting \numselectedpapers\ articles for a detailed review. First, we investigated which data are employed in which areas of cybersecurity training, how, and why. Second, we examined the applications and impact of research in this area, and third, we explored the community of researchers. Our contribution is a systematic literature review of relevant papers and their categorization according to the collected data, analysis methods, and application contexts. These results provide researchers, developers, and educators with an original perspective on this emerging topic. To motivate further research, we identify trends and gaps, propose ideas for future work, and present practical recommendations. Overall, this paper provides in-depth insight into the recently growing research on collecting and analyzing data from hands-on training in security contexts.}

\keywords{cybersecurity education, hands-on training, data science, literature survey, systematic literature review}

\maketitle

\section{Introduction}
\label{sec:intro}

Cybersecurity education occurs at universities, extracurricular events, in the industry, and beyond. In all these cases, cybersecurity training is fundamentally practical. It involves exercising one's skills in applied computing topics, such as administration of operating systems, network attacks and defense, and secure programming. At the same time, cybersecurity is a complex, ever-evolving domain. Instructors and students need to keep up with the latest cyber threat landscape development through hands-on experience.

Because of its hands-on nature, cybersecurity training relies on interactive learning environments and testbeds: technologies that enable students to practice their skills in realistic computer systems. This opens opportunities for collecting objective evidence about learning processes, such as the used commands, system logs, and captured network traffic.

These pieces of evidence collected from cybersecurity training, also called the \textit{training artifacts}, allow researchers to authentically reconstruct the actions that students performed while solving the training tasks. This provides the basis for achieving important educational goals, such as to:
\begin{itemize}
    \item understand students' learning processes and approaches to solving the training assignments;
    \item assess students and measure their learning; and
    \item provide personalized, targeted instruction and feedback.
\end{itemize}

These goals align with the objectives of \textit{educational data mining} (EDM)~\citep{handbook-edm2010} and \textit{learning analytics} (LA)~\citep{handbook-la2017}, two growing disciplines that leverage data from educational contexts. They aim to better understand and improve teaching and learning~\citep{hundhausen2017} by employing computing methods, such as:
\begin{itemize}
    \item mathematical modeling,
    \item automated data analysis,
    \item mining of patterns and processes,
    \item natural language processing, and
    \item machine learning.
\end{itemize}
EDM/LA research findings support more effective training of security professionals, who are direly needed to handle the current cyber threats.

Since educational research is rooted in social sciences~\citep{Malmi2010}, EDM/LA researchers traditionally collected and analyzed data from questionnaires and interviews~\citep{handbook-edm2010, handbook-la2017}. However, these data may not always constitute objective evidence. Students can, intentionally or unintentionally, report behavior or attitudes that differ from the truth. Therefore, it is difficult to ensure the validity and reliability of conclusions drawn from these data~\citep[Chapter 8.2.1]{handbook-edm2010}. That is why we focus on training artifacts measured in interactive learning environments and their application in cybersecurity education research. 

\subsection{Goal of This Paper}

Our goal is to understand the current landscape of EDM/LA in hands-on cybersecurity education. We seek to provide an original overview of this emerging research area that integrates technology and learning. To achieve this goal, we examine the published research on leveraging data from computer systems for cybersecurity training.

Specifically, we perform a systematic literature review (SLR) of research papers that analyze artifacts generated as a product of cybersecurity training. These topics have been recently gaining interest, but they were studied mostly in isolation. We contextualize and organize the research efforts and propose practical implications for further research.

Our findings are relevant in the cybersecurity domain, as well as related areas such as operating systems and networking. They may inspire other researchers who use training environments and collect data from them. Based on our SLR, researchers can learn what data sources to analyze and which approaches were covered in previous work.

\subsection{Research Topics}
\label{subsec:rt}

We aim to examine three research topics in the surveyed papers.
\begin{enumerate}
    \item \textit{Domain and Data.} Since high-quality data are the key to conducting EDM/LA research, we focus on how the researchers addressed the issues of data collection and analysis. First, we explore the cybersecurity areas in which EDM/LA was applied. Then, we investigate how the data were collected, from which computer systems, and categorize the data into distinct types. We also look at the sample size, the time span of the data, how privacy was addressed, and methods for data analysis.
    \item \textit{Research Impact.} Next, we look at whether the methods or findings of the published research were applied in teaching practice, what were the main contributions, and what was the citation impact of the research.
    \item \textit{Research Community.} Finally, we examine the background of researchers in the field, summarize their chosen publication venues, and show how the selected papers are interrelated. Only highlights from this topic are addressed, since this SLR does not primarily focus on bibliometric aspects (such as in~\citep{Zurita2020}).
\end{enumerate}

\Cref{subsec:rq} defines specific research questions within each of these topics. Answers to these questions provide an overview of this novel area for both new and well-established researchers.

\subsection{Paper Structure}

\Cref{sec:related-work} explains the terminology and describes related primary and secondary studies. \Cref{sec:methods} details the methods for conducting the SLR and the review protocol. \Cref{sec:results} presents and discusses the findings. \Cref{sec:implications} proposes ideas for future research and provides guidelines for EDM/LA researchers. Finally, \Cref{sec:conclusion} concludes and summarizes our contributions.

\section{Background and Related Work}
\label{sec:related-work}

This section provides a brief background to cybersecurity training and EDM/LA. It presents related publications and compares them to this paper to explain how we differ from state of the art. \Cref{subsec:glossary} explains the popular formats of cybersecurity training to familiarize readers with the terminology. The overview of the related literature surveys focuses on two domains: \textit{cybersecurity training} (\Cref{subsec:related-work-cyber}) and \textit{educational data analysis} (\Cref{subsec:related-work-edmla}), since this paper is situated in their intersection.

\subsection{Glossary of Hands-on Cybersecurity Training}
\label{subsec:glossary}

In this paper, we consider all hands-on learning sessions during which the students practice their cybersecurity skills. The nomenclature for these sessions varies widely in the literature: they can be called labs, exercises, assignments, or practicals, and they can involve individual or team learning. Below, we particularly introduce two specific types of cybersecurity training: \textit{Capture the Flag} (CTF)~\citep{taylor2017} and \textit{Cyber Defense Exercises} (CDXs)~\citep{Vykopal2017}.

Gamification is popular in many application contexts, including education~\citep{Kasurinen2018, Graham2020}, and CTF is a flagship example of gamifying cybersecurity training. In a CTF game, the trainees solve cybersecurity assignments that yield flags: textual strings that are worth points. There are two main variations of the CTF format: \textit{jeopardy} and \textit{attack-defense}~\citep{Svabensky2021}.

In a jeopardy CTF, trainees choose the tasks from categories such as cryptography, reverse engineering, or forensics. They solve the tasks locally at their computers or interact with a remote server or network. This popular format is hosted even by tech giants like Google in their Google CTF~\citep{googlectf}.

In an attack-defense CTF, teams of trainees each maintain an identical instance of a vulnerable computer system. Each team must protect its system while exploiting vulnerabilities in the systems of other teams. Examples of these events include DEF CON CTF~\citep{defconctf} and iCTF~\citep{vigna2014}.

CDXs such as Locked Shields~\citep{locked-shields}, Crossed Swords~\citep{crossed-swords}, and Cyber Storm~\citep{cyber-storm} aim at professionals, often from military or government agencies or dedicated cybersecurity teams. The trainees form teams whose roles are denoted by colors. \textit{Blue} teams are responsible for maintaining and defending a complex network infrastructure against the attacks of a \textit{red} team. Blue teams must preserve the availability of the network services for end-users. Both CTF and CDXs employ interactive learning environments that allow collecting vast arrays of valuable data, which we explore in this paper.

\subsection{Literature Surveys in Cybersecurity Education}
\label{subsec:related-work-cyber}

The closest paper to ours is a survey by Maennel~\citep{Maennel2020}, who reviewed various data sources that can serve as evidence of learning in cybersecurity exercises. These data sources include timing information, command-line data, counts of events, and input logs. We chose a different methodology (see \Cref{sec:methods}) and posed additional research questions to examine the current literature. Therefore, we provide a complementary and extended perspective.

Švábenský et al.~\citep{Svabensky2020} performed a SLR of 71 cybersecurity education papers published at ACM SIGCSE and ACM ITiCSE conferences since 2010. They investigated which cybersecurity topics were published at these conferences, the teaching context, research methods, citations, and authors within the conferences' community. They found that the examined research primarily employed data from questionnaires and tests to evaluate student perceptions or learning gains. The difference is that this SLR focuses on the applications of EDM/LA and not on any specific venue or time period.

While the review~\citep{Svabensky2020} focused mostly on university education, a review by Khando et al.~\citep{Khando2021} focused on security awareness in organizations. They discovered that various methods, such as gamification and theoretical models, are used to enhance the security awareness of employees. Yet, the paper did not examine the data sources that can be mined in these security awareness programs.

Yamin et al.~\citep{yamin2020} surveyed cyber ranges and security testbeds, platforms that provide technical infrastructure for cybersecurity training. They found that most of these platforms use various data collection mechanisms, including event logging and network layer monitoring. Kucek and Leitner~\citep{Kucek2020}, on the other hand, compared the functionality of open-source environments for conducting CTF sessions. Again, from the data collection perspective, most environments log statistics such as the number of solved challenges and scoring data. However, none of the related papers in this section addressed the research questions we pose in \Cref{subsec:rq}.

\subsection{Literature Surveys in EDM/LA and Computing Education}
\label{subsec:related-work-edmla}

Several literature reviews about EDM/LA were published in the past few years~\citep{linan2015}. The covered topics include evaluation of LA interventions~\citep{Knobbout2020}, LA-driven learning design~\citep{Mangaroska2019}, and LA dashboards~\citep{Matcha2020}. A survey of 240 EDM works~\citep{penaayala2014} identified the most prominent approaches used in EDM research, which include Bayes theorem, decision trees, instances-based learning, and hidden Markov models. There is also a SLR of methods, benefits, and challenges of LA~\citep{avella2016}.

A paper related to ours is a thorough literature review of EDM/LA in programming~\citep{ihantola2015}. It evaluated the content and quality of 76 papers, examining the information that \enquote{can be gained through the analysis of programming data}, and \enquote{which of that data can be collected and analyzed automatically}. These data include keystrokes, line edits, program compilation, program execution, and more. Our paper also focuses on data collection and analysis, however, in cybersecurity training. Moreover, we evaluate the impact and applications of the published research and examine the research community. 

Another thorough survey is by Luxton-Reilly et al.~\citep{Luxton-Reilly:2018}, who reviewed and classified 1666 publications on introductory programming education. The survey highlighted that programming data are used to examine students' compilation behavior, code correctness, and code style. These aspects are studied to identify student competencies and difficulties, predict their performance, and recognize demotivation, among other use cases.

Margulieux et al.~\citep{Margulieux2019} reviewed 197 texts to identify variables measured in computing education papers. These include student performance and information about the timing, progress, and collaboration, for example.

Lastly, Papamitsiou et al.~\citep{Papamitsiou2020} analyzed keywords in 1274 computing education papers to discover clusters of recurring topics. Among the most frequent are assessment, introductory programming, games, and computational thinking.

All these papers indicate a vast potential for EDM/LA in computing education, yet little is known about its application in the field of cybersecurity training. Our article aims to close this gap by examining this emerging topic.

\section{Method of Conducting the Systematic Literature Review}
\label{sec:methods}

To perform this study, we followed the well-established guidelines for conducting a SLR~\citep{prisma2009,kitchenham2007}. We also consulted recommendations for a systematic mapping study~\citep{Petersen:2008, petersen2015} and a literature review section for a Ph.D. dissertation~\citep{randolph2009guide}. This section presents the SLR protocol, which specifies the research questions, search process, and criteria for including the discovered papers.

\subsection{Research Questions}
\label{subsec:rq}

We seek to answer the following research questions to understand the state of the art at the intersection of EDM/LA and cybersecurity training.

\subsubsection*{Research Topic 1: Domain and Data}

\begin{itemize}[leftmargin=1.2cm]
    \item[RQ1.1] In which \textit{areas} of cybersecurity training was EDM/LA applied?
    \item[RQ1.2] What was the \textit{intent} of the data collection?
    \item[RQ1.3] From which computer systems or \textit{environments} were the data collected?
    \item[RQ1.4] What \textit{types of data} were collected from these systems?
    \item[RQ1.5] From \textit{how many students} were the data collected?
    \item[RQ1.6] What was the \textit{time span} of the data? In other words, how long did the educational activity last while the data were collected?
    \item[RQ1.7] Since EDM/LA involves collecting data about people, did the research address data anonymization and \textit{privacy} preservation?
    \item[RQ1.8] Which \textit{analysis methods} were applied to the collected data?
\end{itemize}

\subsubsection*{Research Topic 2: Research Impact}

\begin{itemize}[leftmargin=1.2cm]
    \item[RQ2.1] In which \textit{educational context} was the research practically applied?
    \item[RQ2.2] What were the \textit{contributions} of the research?
    \item[RQ2.3] What were the \textit{supplementary materials} of the research?
    \item[RQ2.4] How much was the research \textit{cited}?
\end{itemize}

\subsubsection*{Research Topic 3: Research Community}

\begin{itemize}[leftmargin=1.2cm]
    \item[RQ3.1] Who were the \textit{authors} of the research, and what were their affiliations?
    \item[RQ3.2] What are the characteristics of the \textit{conferences and journals} they choose for publishing?
    \item[RQ3.3] How much did the members of the community \textit{cite} each other?
\end{itemize}

\subsection{Identifying Sources for the Automated Search for Papers}

We decided not to search for papers in the databases of individual publishers, such as the ACM Digital Library or IEEE Xplore, to avoid inaccuracies and conflicts when merging the results. Instead, we considered three aggregate databases: Web of Science, Scopus, and Google Scholar.

We ultimately used Scopus~\citep{scopus}, since it indexes a representative portion of the databases of individual publishers. We did not choose Web of Science because it does not index several years of relevant educational conferences, such as ACM SIGCSE. We also omitted Google Scholar since it indexes many lower-quality publications, such as non-peer-reviewed papers.

\subsection{Selecting the Keywords for the Automated Search}

When defining the search terms, we aimed to cover the intersection of cybersecurity education and data analysis. We collected the keywords from multiple sources: previously known relevant papers, our expertise, and the knowledge of three cybersecurity experts independent from the paper authors. After multiple iterations and test searches, we established the search query in \Cref{figure:search-query}.

\begin{figure*}[!ht]
    \centering
{\small
\begin{verbatim}
(
  (
    (cybersecurity OR "cyber security" OR "computer security"
      OR "information security" OR "network security")
    AND
    (educat* OR teach* OR instruct* OR student* OR learner OR exercis*)
  )
  OR
  ("capture the flag" OR "cyber defense exercise" OR "cyber defence
    exercise" OR "security training" OR "security exercise" OR "cyber range")
)
AND
(analy* OR evaluat* OR examin*)
\end{verbatim}}
    \caption{The query for the automated search for papers in the Scopus database. Asterisks represent wildcards, and the search is case-insensitive.}
    \label{figure:search-query}
\end{figure*}

After several pilot searches, we excluded the keywords \texttt{learn*} and \texttt{train*}, as they matched hundreds of general machine learning papers about deep learning or training classifiers. It is important to note that this exclusion did not eliminate educational papers. Publications about teaching or learning included at least one of the other educational keywords we used, such as \texttt{educat*}, \texttt{teach*}, or \texttt{student*}. Moreover, we added keywords specific to cybersecurity education, such as \texttt{security training} or \texttt{security exercise}.

We also excluded the keyword \texttt{security}, since it yielded too many irrelevant papers (for example, about fire safety or physical security). Finally, we removed keywords related only to operating systems or networking, since during test searches, the candidate set of results was huge. Nevertheless, the query remained broad enough to avoid the risk of missing a relevant paper.

\subsection{Performing the Automated Search for Candidate Papers}

\begin{figure*}[!ht]
    \centering
    \includegraphics[width=\textwidth]{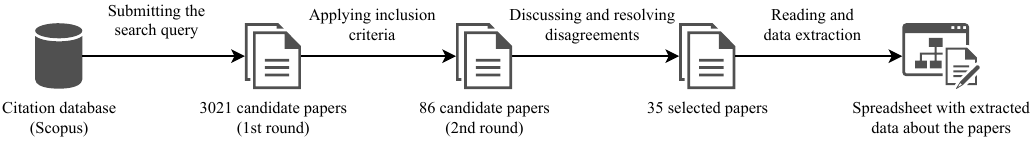}
    \caption{Overview of the steps of the systematic literature review, along with the numbers of papers at each stage.}
    \label{figure:overview}
\end{figure*}

\Cref{figure:overview} shows an overview of the SLR process. We started by submitting the query in \Cref{figure:search-query} to the online database Scopus~\citep{scopus}. We restricted the search to titles, abstracts, and keywords of papers in conference proceedings or journals, in the area of computer science or engineering, and in the English language. Then, we exported the results as bibliographic records (in the \texttt{bib} format) to the Mendeley reference manager~\citep{mendeley}.

We performed the first search on March 29, 2019. Immediately afterward, we subscribed to Scopus e-mail notifications that informed us about newly indexed papers, which we gradually added to the candidate set. We stopped adding new candidates on \datesearchended.

This process yielded \numtotalpapers\ candidate papers. To evaluate the search, we checked that the candidate set included relevant papers that we knew from our previous research in cybersecurity education.

\subsection{Defining the Inclusion and Exclusion Criteria}

After multiple iterations and pilot tests, we defined five selection criteria. We explain the rationale behind each criterion and provide examples of exclusion when necessary.

\begin{enumerate}
    \item The paper must have \textit{full text} available and be at least \textit{four pages} long. Shorter papers lacked space to provide detailed information needed to answer our research questions.\smallskip
    
    \item The paper must \textit{report on collected data} that originate from \textit{human interaction} with a \textit{computer system} for cybersecurity training. By a computer system, we mean a hardware, or virtualized, or cloud infrastructure, in which the student interacts with software applications. This comprises either \textit{individual} or \textit{team} learning that occurs during labs, training sessions, simulations, competitions, exercises, and the like. \smallskip
    
    We excluded papers that dealt with video and board games. Although these games have educational potential, they do not emulate realistic cybersecurity operations that involve cybersecurity tools and their monitoring. We also excluded purely system design papers that only proposed a data collection/analysis toolchain but did not apply it in practice, not even in author testing. The reasoning from our experience was that the system might theoretically look good as a proposal on paper, but its true capabilities can only be proven in practice. So, our research questions focus on practical demonstration and application.\smallskip
    
    \item The collected data must result either from interactions with an \textit{operating system} and \textit{applications} running in it (for example, recorded keystrokes, mouse clicks, memory dump, filesystem changes, or network traffic) or with some additional \textit{training system} (such as timings of actions in a separate training interface).\smallskip
    
    This means we excluded papers that reported only grades of students or only administered questionnaires (for example,~\citep{chothia2015offline}). As we argued in \Cref{sec:intro}, we consider these data often not representative of students' actions. For similar reasons, we also excluded papers focusing solely on the affective domain, such as emotion recognition~\citep{Imani2019}. \smallskip
    
    \item The data must be collected and analyzed \textit{automatically} or at least semi-automatically. We excluded papers that involved only fully manual data processing, such as human graders observing the students and noting their actions (for example,~\citep{Rege2017}). This approach is time-consuming to replicate, does not scale, and is prone to errors. Therefore, our review targets computer technologies for automated data processing.\smallskip
    
    \item The data analysis must support an \textit{educational goal}, for example, to assess students or help instructors understand the students' learning processes or behavior. This means we excluded papers that presented only performance testing of the learning environment.
\end{enumerate}

\subsection{Preliminary Reading and Applying the Inclusion Criteria}

Two authors, independently of each other, preliminarily screened each of the \numtotalpapers\ candidate papers and applied the inclusion criteria. We followed the process in \Cref{figure:selection-algorithm}.

\begin{figure}[!ht]
    \centering
\begin{lstlisting}
    |\textbf{for}| each paper in the candidate set:
        read the title and abstract
        decide for inclusion or exclusion
        |\textbf{if}| decision cannot be made:
            read the introduction and conclusion
            decide for inclusion or exclusion
            |\textbf{if}| decision cannot be made:
                skim-read the rest of the paper
                decide for inclusion or exclusion
\end{lstlisting}
    \caption{The algorithm for selecting papers for the literature review from the candidate papers.}
    \label{figure:selection-algorithm}
\end{figure}

\subsection{Resolving Disagreements and Selecting Papers for Review}

When the two readers finished, they compared their decisions. There were three possibilities:
\begin{itemize}
    \item If both readers voted to include the paper, it was selected.
    \item If both readers voted to exclude the paper, it was rejected.
    \item If there was a disagreement, it was resolved by discussion. Afterward, the paper was either selected or rejected.
\end{itemize}

The readers initially agreed on 97.8\% of papers (23 immediate selections and 2935 immediate rejections). We initially disagreed and further discussed the remaining 2.2\% (63 papers). Our inter-rater agreement~\citep{krippendorff2004reliability} measured by Scott's $\pi$ and Krippendorff's $\alpha$ (for nominal data) was 0.41, which is moderate. The coefficients were calculated using the Python NLTK module~\citep{nltk}.

\subsection{Extracting Data from Selected Papers}

We selected \numselectedpapers\ papers for the detailed review. We read their full texts, extracted the information determined by our research questions, and recorded them in a spreadsheet. \Cref{sec:results} presents the results.

\section{Results and Discussion}
\label{sec:results}

Although the oldest candidate paper is from the year 1976, the oldest selected paper is from 2012. Moreover, more than two-thirds of the selected papers were published in 2017 or later. This implies that EDM/LA in cybersecurity is an arising topic that has recently started gaining traction, and it will likely continue in this growing trend.

We now answer the research questions from \Cref{subsec:rq}. Throughout this section, we refer to \Cref{table:results-topic-goal} and \Cref{table:results-data}, which summarize the results.

\begin{landscape}

\begin{table*}[!ht]
\caption{Overview of the goals of the \numselectedpapers\ reviewed papers grouped by topics. The papers are identified by an arbitrary number based on the year of publication. For two papers P$x$ and P$y$, if $x<y$, then P$x$ was published before P$y$ or in the same year.}
\tiny
\centering
\rowcolors{2}{gray!10}{white}
\renewcommand*{\arraystretch}{1.2}
\begin{tabular}{lll}
\textbf{Paper ID} & \textbf{Cybersecurity Topics (RQ1.1)} & \textbf{Goal of the Paper} \\
\hline
P3~\cite{chapman2014picoctf} & offense, forensics & analyze the preferences, activity, and number of learners in a CTF \\
P5~\cite{Burket2015problemgeneration} & offense & detect cheating in a CTF by analyzing sharing of solutions \\
P7~\cite{Weiss2016reflective} & offense & assess learners by visualizing their command history as a directed graph \\
P9~\cite{Vykopal2016} & offense & determine what information can be predicted from logs of 260 trainees \\
P11~\cite{Tseng2017ontology} & offense & analyze learners' behavior in a CTF to reveal their misconceptions \\
P12~\cite{caliskan2017capability} & offense & determine metrics from exercise logs that will predict students' grade \\
P14~\cite{Andreatos2017} & offense & analyze students' network activity in a lab to review their actions \\
P16~\cite{Kont2017} & offense & provide and evaluate feedback for the attacking teams in a CDX \\
P18~\cite{Chothia2017jail} & offense & determine if storyline in a cybersecurity training improves learning \\
P19~\cite{Tian2018} & offense & provide trainees with situational awareness of the training \\
P21~\cite{Svabensky2018challenges} & offense & determine if trainees fulfill prerequisites of security training \\
P22~\cite{Svabensky2018gathering} & offense & analyze how trainees interact with security training tasks and tools \\
P25~\cite{Andreolini2019} & offense & assess trainees by comparing their actions to a reference solution \\
P26~\cite{Falah2019} & offense & estimate the difficulty of attacks and measure skills of trainees \\
P28~\cite{Maennel2019} & offense, network security & assess students who apply to a cybersecurity master degree program \\
P31~\cite{Tobarra2020-game} & offense, forensics, network security & compare assessment of students who did / did not participate in a CTF \\
P34~\cite{Kaneko2020} & offense, forensics & evaluate an intensive cybersecurity course based on student performance \\
P35~\cite{Yett2020} & offense, secure programming & analyze how students collaborate in group programming tasks \\
P2~\cite{reed2013instrumenting} & forensics & analyze score distribution, submission delay, and frustration in a CTF \\
P6~\cite{Abbott2015log} & forensics & quantitatively analyze student actions and performance in security training \\
P1~\cite{rupp2012putting} & network security & identify skill profiles of students based on logs and submitted commands \\
P23~\cite{Zeng2018improving} & network security & compare student grades with the time they spent working on lab tasks \\
P24~\cite{Deng2018personalized} & network security & adapt instruction to trainees' learning style, predict their performance \\
P27~\cite{Palmer2019} & network security & automatically assess the quality of students' network configuration \\
P30~\cite{Sheng2020} & network security & evaluate a custom machine learning model for assessing students \\
P33~\cite{Tobarra2020-acceptance} & network security & evaluate how often and how long students interact with a training platform \\
P8~\cite{granaasen2016measuring} & incident response & assess performance, behavior, and progress of teams in a CDX \\
P10~\cite{Henshel2016predicting} & incident response & determine proficiency metrics to assess performance of teams in a CDX \\
P15~\cite{Labuschagne2017} & incident response & assess trainees by comparing their command history with an ideal solution \\
P17~\cite{Maennel2017} & incident response & propose and apply a methodology for measuring learning in a CDX \\
P20~\cite{Kokkonen2018} & incident response & analyze communication patterns of defending teams in a CDX \\
P13~\cite{Weiss2017} & system administration & assess learners by visualizing their command history as a directed graph \\
P4~\cite{Nadeem2015method} & secure programming & recommend reading to developers based on vulnerabilities in their code \\
P29~\cite{Gasiba2020} & secure programming & compare two methods for measuring time to solve a challenge \\
P32~\cite{Almansoori2020} & secure programming & analyze how students and instructors use unsafe C/C++ functions \\
\hline
\end{tabular}
\label{table:results-topic-goal}
\end{table*}

\begin{table*}[!ht]
\caption{Overview of the intent of data collection, collected data, analysis methods, and contributions of the \numselectedpapers\ selected papers. 
}
\tiny
\centering
\rowcolors{2}{gray!10}{white}
\renewcommand*{\arraystretch}{1.2}
\begin{tabular}{lllll}
\textbf{Paper ID} & \textbf{Intent (RQ1.2)} & \textbf{Collected Data (RQ1.4)} & \textbf{Analysis (RQ1.8)} & \textbf{Contribution (RQ2.2)} \\
\hline
P1~\cite{rupp2012putting} & assess & C, T & DS, ML & study of learning in a training platform \\
P2~\cite{reed2013instrumenting} & assess, inform & E, T & DS, NM & measure of situational awareness \\
P3~\cite{chapman2014picoctf} & assess, inform & E, T & DS & open-source platform and challenges \\
P4~\cite{Nadeem2015method} & support & C & DS, NM & architecture of a learning system \\
P5~\cite{Burket2015problemgeneration} & assess & E, T & DS, QA & open-source platform and challenges \\
P6~\cite{Abbott2015log} & assess & H, A, N, E & DS & architecture of data collection infrastructure \\
P7~\cite{Weiss2016reflective} & assess, support & C & QA & study of the utility of command history \\
P8~\cite{granaasen2016measuring} & assess & A, N, I, V & DS, NM & assessment model for defense exercises \\
P9~\cite{Vykopal2016} & assess, inform & E, T & DS & study of the utility of game logs \\
P10~\cite{Henshel2016predicting} & assess & N, I, T & DS, ML & assessment model \\
P11~\cite{Tseng2017ontology} & assess & H & QA & study of students' behavior \\
P12~\cite{caliskan2017capability} & assess & H, N & DS, ML & study of grading students \\
P13~\cite{Weiss2017} & assess, support & C & QA & study of the utility of command history \\
P14~\cite{Andreatos2017} & assess & N & DS & study of monitoring student network activity \\
P15~\cite{Labuschagne2017} & assess & C, T & DS, NM & method for scoring \\
P16~\cite{Kont2017} & support & H, N & DS & framework for attackers' situational awareness \\
P17~\cite{Maennel2017} & assess & N, T & QA & method for measuring learning \\
P18~\cite{Chothia2017jail} & assess & C & DS & study of story improving engagement \\
P19~\cite{Tian2018} & assess, support & H, N, D, C & QA & study of the utility of command history \\
P20~\cite{Kokkonen2018} & assess & I, T & QA & tool for CDX organizers' situational awareness \\
P21~\cite{Svabensky2018challenges} & assess & E, T & DS, ML, QA & study of predicting prerequisites \\
P22~\cite{Svabensky2018gathering} & assess, inform & E, T & DS, QA & study of students' behavior \\
P23~\cite{Zeng2018improving} & assess & T & DS & study of factors that contribute to learning \\
P24~\cite{Deng2018personalized} & assess, support & C, T & DS, ML & method for personalizing instruction \\
P25~\cite{Andreolini2019} & assess & H, A, N, C, T & DS, NM, QA & method for modeling and scoring training \\
P26~\cite{Falah2019} & assess & E, T & DS, PM & method for scoring / assessing performance \\
P27~\cite{Palmer2019} & assess & D & DS & tool for assessing performance \\
P28~\cite{Maennel2019} & assess & E, T & QA & lessons learned from assessing students \\
P29~\cite{Gasiba2020} & assess, support, inform & E, T & DS, NM, PM, QA & methods for computing challenge solve time \\
P30~\cite{Sheng2020} & assess & N & NM, ML & experimental comparison of two metrics \\
P31~\cite{Tobarra2020-game} & assess & E, T & DS & study of effect of CTF on grades \\
P32~\cite{Almansoori2020} & support & C & DS, NM & study of issues in C/C++ code at universities \\
P33~\cite{Tobarra2020-acceptance} & inform & E, T & DS, IS & study of students' interactions with a platform \\
P34~\cite{Kaneko2020} & assess & H, A, V, T & QA & study of an exercise-based cybersecurity course \\
P35~\cite{Yett2020} & assess & E, T & DS, IS, PM, QA & study of students' approaches to collaborative tasks \\
\hline
\end{tabular}
\label{table:results-data}

\medskip

\textit{Collected data}: C = shell commands and program code, N = network logs and traces, H = host-based logs, A = application logs, D = disk and memory content, E = training events, I = interaction and communication, V = video, T = timestamps.

\textit{Analysis methods}: DS = descriptive statistics, IS = inferential statistics, NM = numerical methods, ML = machine learning, PM = probabilistic modeling, QA = qualitative analysis.
\end{table*}

\end{landscape}

\subsection{Research Topic 1: Domain and Data}

We start by looking at the first eight research questions about the data collection and analysis.

\subsubsection{RQ1.1: Cybersecurity Topics}

We categorize the papers into custom technical topics and also identify the topics from the CSEC2017 cybersecurity curriculum~\citep{cybered}. The categorization intentionally omits soft skills such as critical thinking and teamwork, which are outside the scope of this SLR.

As \Cref{table:results-topic-goal} shows, 18 papers focus on teaching offensive security skills, including penetration testing, exploitation, network attacks, cryptographic attacks, and reverse engineering. 22 papers focused on defensive skills, which we divided into the following:
\begin{itemize}
    \item \textit{Network security} (P1, P23, P24, P27, P28, P30, P31, P33), which includes technical defensive skills, such as configuring networks, firewalls, and intrusion detection.
    \item \textit{Incident response} (P8, P10, P15, P17, P20), which involves the network security skills applied while resolving a simulated cybersecurity incident.
    \item \textit{Forensic analysis} and examining digital evidence (P2, P3, P6, P31, P34).
    \item \textit{Secure programming} and preventing vulnerabilities (P4, P29, P32, P35).
    \item \textit{System administration} (P13), which involves configuring a Linux system.
\end{itemize}

Next, we mapped the topics onto the Knowledge Areas of the CSEC2017 curriculum~\citep{cybered}. The mapping revealed that \textit{Connection security} and \textit{System security} are dominantly represented (in 26 and 23 papers, respectively). Also present were \textit{Data security} (7 papers), \textit{Software security} (4 papers), and \textit{Component security} (3 papers). \textit{Human}, \textit{Organizational}, and \textit{Societal security} were not present due to our inclusion criteria. Interestingly, although programming topics are prevalent in computing education research, there were only four papers on secure programming in our dataset.

However, the topic mapping was sometimes difficult. Only a minority of papers stated the learning objectives or described the cybersecurity skills they aim to practice. For example, in P2, we were unsure about the content of the exercises. We assigned the paper in the Forensics category because it stated that \enquote{The challenges contained forensics data}. Moreover, very few papers referenced a standardized cybersecurity curriculum, such as~\citep{cc2020, sahami2013curricula}, when defining their learning objectives. Nevertheless, we managed to work with the information that was apparent from the paper text.

\subsubsection{RQ1.2: Intent of Data Collection}

We synthesized three main purposes of data collection:
\begin{itemize}
    \item \textit{Assess students}, that is, measure their performance, reveal their misconceptions, analyze their task solution patterns, or otherwise evaluate their actions. This was the goal of 31 papers, the vast majority.
    \item \textit{Support learning} of students or provide feedback to them. This was the goal of 8 papers. Some publications had multiple goals, so 5 of these papers overlapped with the assessment papers.
    \item \textit{Inform the training content creators} to provide them feedback about how the students approached the tasks. This was the goal of 6 papers.
\end{itemize}

\Cref{table:results-data} categorizes the selected papers based on the intent. Although several papers have the same overarching goal, they achieve it with different applications of EDM/LA. We gradually analyze various aspects of these applications in the following sections. 

\subsubsection{RQ1.3: Environments for Data Collection}

A learning environment that allows automated data collection is a crucial prerequisite for any EDM/LA applications. Therefore, it was one of the aspects on which we focused in our review. Specifically, we observed four types of environments from which the data were collected:
\begin{itemize}
    \item The \textit{training infrastructure}, which is a physical or virtual environment that consists of one or more hosts with a standard operating system. The hosts are usually networked. This category includes cyber ranges~\citep{yamin2020} and lab platforms, and 24 papers collected data from them.
    \item A \textit{software application} that simulates a network environment. This was applicable only for P1 and P27.
    \item \textit{Learning management system} (LMS), which is a web-based technology that facilitates the training, such as a CTF platform~\citep{Kucek2020}. LMS allows collecting the solutions to tasks submitted by students. This applied to 8 papers.
    \item \textit{External sources} of data. This was applicable only for P4 and P32 that collected source code from repositories.
\end{itemize}

\subsubsection{RQ1.4: Collected Data}

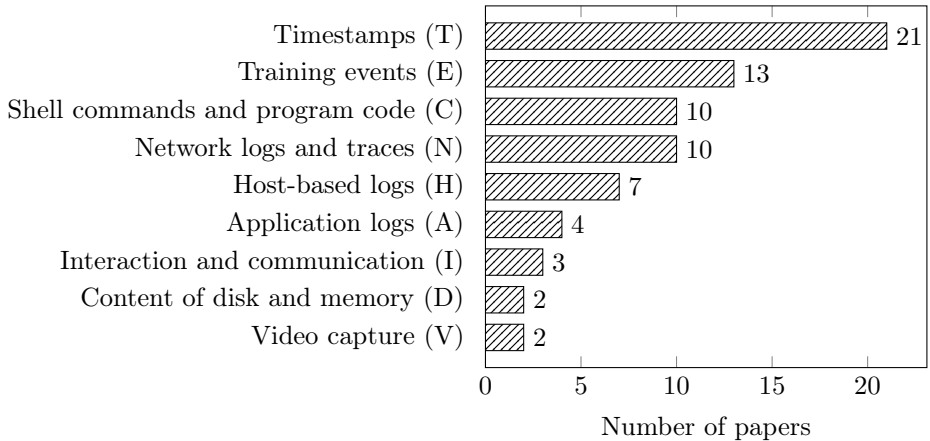
\begin{figure*}[!ht]
    \centering
\begin{tikzpicture}
  \begin{axis}[
    xbar,
    xmin=0,
    width=7.4cm, height=6.4cm, enlarge y limits=0.1,
    xlabel={Number of papers},
    symbolic y coords={Video capture {(V)}, Content of disk and memory {(D)}, Interaction and communication {(I)}, Application logs {(A)}, Host-based logs {(H)}, Network logs and traces {(N)}, Shell commands and program code {(C)}, Training events {(E)}, Timestamps {(T)}},
    ytick=data,
    ytick style={opacity=0},
    nodes near coords, nodes near coords align={horizontal},
    ]
    \addplot[black,pattern=north east lines] coordinates {(2,Video capture {(V)}) (2,Content of disk and memory {(D)}) (3,Interaction and communication {(I)}) (4,Application logs {(A)}) (7,Host-based logs {(H)}) (10,Network logs and traces {(N)}) (10,Shell commands and program code {(C)}) (13,Training events {(E)}) (21,Timestamps {(T)})};
  \end{axis}
\end{tikzpicture}
    \caption{The distribution of how often the identified data types were present in the \numselectedpapers\ papers.}
    \label{figure:bar-chart-data-types}
\end{figure*}

The data collected from learning contexts were largely heterogeneous, demonstrating the diverse possibilities that EDM/LA offers. We synthesized nine categories denoted by capital letters used in~\Cref{table:results-data} and \Cref{figure:bar-chart-data-types}.

The following data were collected from the training infrastructure or simulation software as a result of the students' direct interaction with the training environment:
\begin{itemize}
    \item \textit{Shell commands and program code} (C), including that from external sources (10 papers).
    \item \textit{Network logs and traces} (N), including packet captures and intrusion detection system logs (10 papers).
    \item \textit{Host-based logs} (H), which include Syslog, audit logs, event logs, CPU and memory usage, and process activity (7 papers).
    \item \textit{Application logs} (A), such as the status of network services, interaction with a graphical user interface, mouse clicks, and keystrokes (4 papers).
    \item \textit{Content of disk and memory} (D), such as stored configuration files (2 papers).
\end{itemize}

The data generated during the training collected from LMS and other sources comprise these categories:
\begin{itemize}
    \item \textit{Training events} (E), that is, interactions with a LMS or actions such as submitted answers or flags in a CTF (13 papers)\footnote{For an in-depth overview of learning events and architecture for collecting them, see~\citep{Ayres2017}.}.
    \item \textit{Interaction and communication} (I) between students, which includes chat content or e-mail headers (3 papers).
    \item \textit{Video capture} (V) of learners' screens (2 papers).
\end{itemize}

The final category of data collected in 21 papers was \textit{timestamps} (T) of actions such as command submissions or event triggers. This category also included the corresponding derived data, such as the duration of these actions.

In the reviewed papers, we considered only data that were collected in practice and later analyzed. For example, P15 states that their cyber range allows collecting host-based and network logs. However, the paper does not demonstrate this capability in practice. The analysis is performed only on shell commands and time-related information, so the corresponding entry for P15 in~\Cref{table:results-data} lists only the categories C and T.

Apart from the data relevant to our research questions, some papers employed other data sources, for example, learner surveys (P8, P10, P21, P22, P24, P28, P31, P33), student grades (P12, P18, P23), observer reports (P8), sample solutions to tasks (P15, P25), and Common Weakness Enumeration (CWE) articles (P4, P29). For more additional data types, we refer the reader to the above-mentioned survey by Maennel~\citep{Maennel2020}.

\subsubsection{RQ1.5: Sample Size}

The size of the collected dataset influences which EDM/LA methods are applicable. For example, many machine learning techniques require thousands or tens of thousands of data points. However, due to the diversity of data types we identified, it is impossible to compare the papers directly. For example, a dataset of 1 GB of network traffic and 1 hour of video footage of learners' screens are incomparable.

Therefore, as a proxy, we looked at the number of participants from which data were collected. The sample size ranged widely, from one (P19) to 9738 (P3, P5) participants. The median was 43. For comparison, in general cybersecurity education papers, the median number of participants is about 40~\citep{Svabensky2020}.

Although most papers reported the number of participants, there were occasional issues with clarity. For example, P17 states that 900 people participated in the exercise; however, it was unclear whether all of them contributed to the dataset. In P7, 24 teams participated, but the team size is unknown\footnote{When computing the median of participants, we performed a small simplification for P7: by assuming two or three people per team, we estimated 60 participants in the 24 teams.}.

\subsubsection{RQ1.6: Time Span of Data}

We also looked at the time span during which the data were collected since this is another proxy indicator of the dataset's depth. Most commonly, 14 papers collected the data during the period from 1 to 14 days. This was usually the case of CTF and CDX. Next, ten papers collected the data over a period from 1 to 13 hours. They mostly examined one or more lab sessions. Three papers spanned a month or more. Finally, eight papers did not report the time-related information.

\subsubsection{RQ1.7: Privacy and Ethical Issues}

Although almost all papers collected data about human participants, only eight publications explicitly addressed privacy and ethical issues. P3 explains that no information about individual students was recorded. Similarly, P22 and P28 explain that the collected data were not linked to personally identifiable information during the research. In P21, the data were anonymized, and P17 argues that only aggregate data are presented to preserve anonymity. Finally, P6 and P29 describe that participants explicitly consented to data collection, and in P8, the participants could opt out of the data collection.

Ethical measures, such as data anonymization, may be overlooked when reporting EDM/LA research. However, they constitute an important part of the research process, so EDM/LA researchers should not neglect them in future work.

\subsubsection{RQ1.8: Analysis Methods}

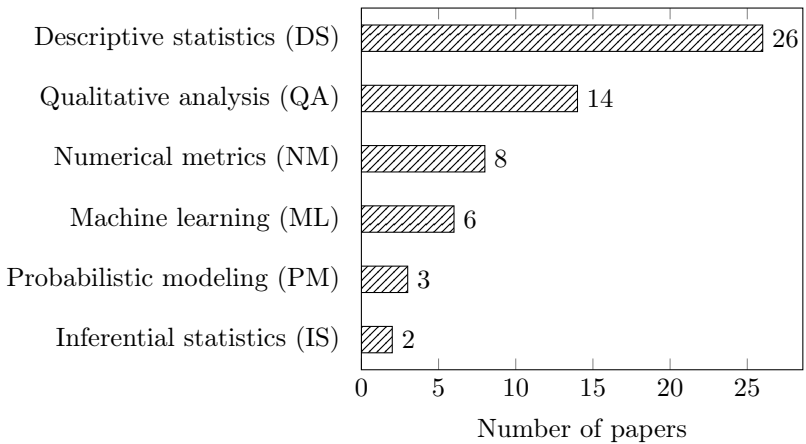
\begin{figure*}[!ht]
    \centering
\begin{tikzpicture}
  \begin{axis}[
    xbar,
    xmin=0,
    width=7.4cm, height=6.4cm, enlarge y limits=0.1,
    xlabel={Number of papers},
    symbolic y coords={Inferential statistics {(IS)}, Probabilistic modeling {(PM)}, Machine learning {(ML)}, Numerical metrics {(NM)}, Qualitative analysis {(QA)}, Descriptive statistics {(DS)}},
    ytick=data,
    ytick style={opacity=0},
    nodes near coords, nodes near coords align={horizontal},
    ]
    \addplot[black,pattern=north east lines] coordinates {(2,Inferential statistics {(IS)}) (3,Probabilistic modeling {(PM)}) (6,Machine learning {(ML)}) (8,Numerical metrics {(NM)}) (14,Qualitative analysis {(QA)}) (26,Descriptive statistics {(DS)})};
  \end{axis}
\end{tikzpicture}
    \caption{The distribution of how often the identified analysis methods were present in the \numselectedpapers\ papers.}
    \label{figure:bar-chart-analysis-methods}
\end{figure*}

EDM/LA offer a multitude of techniques and methods for the analysis of collected data to achieve an educational goal. We observed these six types of analysis, which are summarized in~\Cref{table:results-data} and \Cref{figure:bar-chart-analysis-methods}.

\begin{itemize}
    \item \textit{Descriptive statistics} (DS) of collected data, including their correlations. Almost all papers (26 out of \numselectedpapers) reported some descriptive statistics.
    \item \textit{Inferential statistics} (IS), which was present only in P33 and P35. Although P8, for example, applied statistical testing to questionnaire data, these data were out of the scope of this SLR. Nevertheless, since the median sample across all selected papers was 43 participants, many statistical tests might not have been appropriate in other papers.
    \item \textit{Numerical metrics} (NM) computed by aggregating the collected data, usually for assessment. Eight papers employed standard metrics, such as cosine similarity, or developed custom scoring metrics.
    \item \textit{Machine learning} (ML) methods, more specifically:
    \begin{itemize}
        \item regression (P10, P21);
        \item classification using Naive Bayes (P12), decision trees (P12, P24), or support vector machines (P24);
        \item principal components analysis (P1); and
        \item custom models (P30).
    \end{itemize}
    \item \textit{Probabilistic modeling} (PM) in P26, P29, and P35.
    \item \textit{Qualitative analysis} (QA) of the collected data, either of the submitted commands (P7, P13, P19) or other student actions (P11, P17, P20, P28, P34). Moreover, six other papers (P5, P21, P22, P25, P29, P35) used qualitative analysis in addition to other methods.
\end{itemize}

Most data analyses were performed after the training ended, which seems to be the most straightforward use case. The exceptions included live scoring during a CDX (for example, P8 and P16).

\subsection{Research Topic 2: Research Impact}

Next, we reviewed the papers' application domain, contribution, supplementary materials, and citation impact.

\subsubsection{RQ2.1: Application in Practice}

The published research had diverse application contexts, which again demonstrates that EDM/LA is suitable for various use cases. 13 papers were applied within university courses. Next, five papers were from a CDX and five more from a jeopardy CTF. Ten were applied in other types of cybersecurity training. Finally, two papers had no application besides the author testing.

\subsubsection{RQ2.2: Contribution}

\Cref{table:results-data} shows an overview of the contribution of the individual papers. In this section, we discuss the specific contribution of each paper in more detail. The papers are grouped by the application context.

The majority of papers used student interaction data within a university course. P7 and P13 generated graphical progress models of students' approaches to solving cybersecurity exercises using a command line. P12 and P14 used mainly network logs to assess students in cybersecurity courses. P18 analyzed whether introducing an optional story element into cybersecurity assignments improves student performance. P23 and P24 assessed student learning mainly based on timing information, such as time spent on the tasks. Similarly, P26, P28, P31, and P33 measured skills of students using time-base statistics and event logs, in addition. P27 evaluated students' network configuration and P32 students' usage of unsafe programming functions.

A compact cluster of papers concentrated on CDX. P8 collected network and system logs to study the performance of participating teams. Similar data sources were used in P10 to assess and predict team performance. P17 proposed a more systematic approach: a methodology to employ CDX data for team assessment. P16 focused on using CDX data to provide feedback for Red teams. Finally, P20 developed a tool to analyze Blue team communication and reporting data.

Jeopardy CTF was another important application area. P2 collected learner interaction data from a CTF platform to measure score distribution, time delays, and frustration of participants. P6 followed up on this work to derive meaningful blocks of learner activity, such as the used applications, from CTF logs. P3 analyzed challenge completion in a large-scale online Jeopardy CTF. P5 used the same dataset as P3 to report on observed cheating attempts and proposed a method called automatic problem generation as a solution. Finally, P11 used host-based logs to describe learner activity.

Most of the remaining works focused on other types of cybersecurity training. P1 proposed an evidence model for analyzing data from Packet Tracer, software for learning networking, to discover skill levels of learners. P9, P21, and P22 evaluated data capturing learner interactions with a cyber range to understand how students approach solving cybersecurity exercises. P15 employed metrics such as timing, commands entered, and similarity to the reference solution to assess technical skills of trainees in a cyber range. P25 generated visual models of trainees' approaches and investigated their difference from the reference solution, again using this information for skill assessment. P29 and P30 created sophisticated numerical and machine learning models to analyze log data from cybersecurity training. P34 used operation logs to evaluate a cybersecurity course, and P35 focused on student collaboration in group programming tasks.

Finally, two papers presented only a prototype. P4 proposed a method for analyzing program code to discover vulnerabilities and recommend relevant sources to software developers. Lastly, P19 proposed a method for real-time analysis of log data from cyber ranges to improve situational awareness.

In a few cases, however, the contribution was difficult to determine. For example, the abstract of P11 states that the data analysis will be used to reveal student misconceptions. However, no reported results indicated the misconceptions were found.

\subsubsection{RQ2.3: Supplementary Materials}

Only eight publications provided supplementary materials along with the paper. Papers about the PicoCTF platform (P3, P5) released the open-source code of the platform on GitHub. The Frankenstack framework (P16) can also be found on GitHub, although the repository is not linked from the paper. Similarly, P21 links a repository with an open-source visualization tool, and P29 refers to open-source components of the training platform. P18 provides virtual machines and cybersecurity exercises for other instructors. P28 also links exercises, but the link is no longer functional. Finally, P33 provides a video about the training platform and solving exercise tasks in it.

\subsubsection{RQ2.4: Citation Analysis}

The citation analysis was conducted on \datecitationanalysis\ using Scopus. Although the citation counts are relatively small (min = 0, max = 64, median = 5), this is probably because 71\% of the papers were published in 2017 or later. As a result, there was not enough time for the citation impact to appear. However, the sample is too small for conclusive results.

Interestingly, although the most cited journal paper (P19) has 64 citations, the number drops to 32 after removing self-citations. On the other hand, the most cited conference paper (P3) has 63 total citations and 62 non-self-citations. This paper deals with PicoCTF, a popular event held annually since 2013, along with an associated open-source platform.

\subsection{Research Topic 3: Research Community}

We now present the results regarding the authors, their affiliations, and publication venues.

\subsubsection{RQ3.1: Authors and Their Affiliations}

A total of 125 unique authors wrote the surveyed papers.
Out of these authors, 101 co-authored only one paper, 21 co-authored two papers, and 2 authors three papers. This suggests that the community of EDM/LA researchers in cybersecurity is neither stable nor particularly big.

Considering the authors' affiliations, 92 of them were associated with a university or a college, 18 with a military or government institution, 13 with a private company, and 4 with a non-governmental research and development organization\footnote{The counts sum to 127 because two authors had two affiliations.}. The prevalence of academic institutions is motivated by the fact that the authors often work as teachers of cybersecurity courses, and the research supports their teaching. Another contributing factor is that their institutions may require them to publish as a part of their job duties.

Of the \numselectedpapers\ selected papers, 22 were written solely by university researchers and 3 by military/government institutions. We observed little cross-institutional collaboration. Universities and military collaborated in 3 cases, universities and private companies in 3 cases, and research institutes and private companies in 4 papers. Such collaboration can be beneficial, because the research addresses the needs of various stakeholders, and the educational intervention is evaluated at multiple institutions.


\subsubsection{RQ3.2: Publication Venues}

The selected papers were published in various conferences and journals; there were no prominent flagship venues for EDM/LA research in cybersecurity. However, some trends appeared: conferences are preferred to journals (27 vs. 8), probably due to the speed of publication and targeting a specific audience. Also, bibliometrics is not an important criterion for most authors. Half of the journal papers were not indexed in Web of Science~\citep{jcr}, and most conferences were not CORE-ranked~\citep{core}. However, other metrics and standards for rating the quality of publication venues also exist, so these provide only a partial point of view.

\subsubsection{RQ3.3: Citation Map}

\begin{figure*}[!ht]
    \centering
    \includegraphics[width=0.85\textwidth]{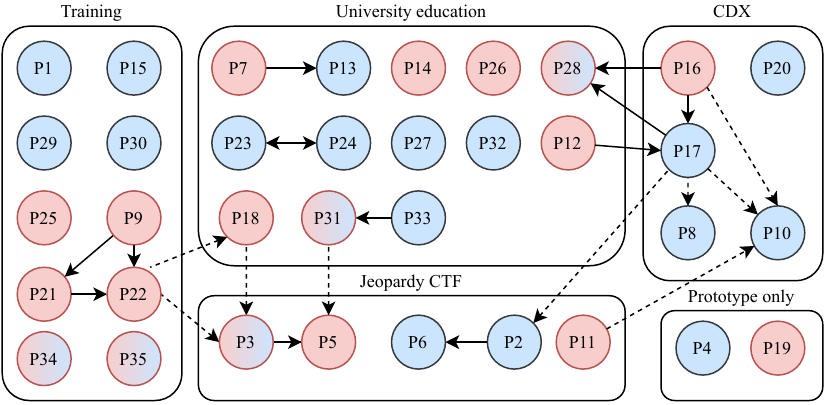}
    \caption{The \numselectedpapers\ papers grouped by the application context (RQ2.1). Red nodes deal with offensive topics and blue nodes with defensive (RQ1.1). A solid arrow indicates overlapping authors and time progression ($A \rightarrow B$ means that $B$ was published after $A$ and shared one or more authors). A dashed arrow indicates non-self citations ($B \dashrightarrow A$ means that $B$ cites a previously published paper $A$ and both have different authors).}
    \label{figure:graph}
\end{figure*}

We examined whether there are citation interconnections between papers that might indicate relationships between researchers. However, as \Cref{figure:graph} shows, the papers rarely cite each other. The most cited paper is P10 with three non-self citations from P11, P16, and P17, indicating that P10 may represent important prior work. Next, P3 has two citations by P18 and P22, both of which deal with offensive cybersecurity topics similar to P3. P5 is cited by P31, again overlapping in topics. Lastly, P2 and P8 have a single citation each, both from P17, focusing on defensive topics. Overall, these links are relatively weak, showing that the community is fragmented, with the only prominent group forming in the CDX application domain.

\subsection{Summary of the Observed Trends}

We now summarize the key results of the literature review. The papers cover offensive and defensive topics almost equally, which can also be seen in \Cref{figure:graph}. The authors are usually from universities or military institutions, and their typical goal is to assess students. To do so, the researchers collect data from the training infrastructure, which is often virtualized.

When sorted by the most common data type, the data include timestamped actions in a learning infrastructure, training events, commands and code, network traces, and host-based logs. They are collected from a median of 43 students over a period of a few hours to several days. To analyze the data, the authors usually present descriptive statistics or qualitative insights.

The research often provides valuable contributions, such as innovative scoring methods, insights into students' behavior, or case studies carried out in an authentic context. As \Cref{figure:graph} shows, the application context is relatively evenly distributed among university education, CDX, CTF, and other types of cybersecurity training.

The papers' main disadvantage is that they rarely provide supplementary materials, preventing other researchers and educators from building on the results. This is also true in other cybersecurity education papers~\citep{Svabensky2020}. Moreover, few of them address the privacy of the students and data anonymization issues.

\subsection{Limitations of the Results}

All literature reviews are limited by the selection of paper databases and search terms. Although we focused only on papers indexed by Scopus, it is a major database, and our search query was broad and reviewed by independent cybersecurity experts. Therefore, we believe we minimized the number of missed candidate papers.

Another limitation of all surveys is a potential researcher bias when manually reviewing the papers and extracting data from them~\citep{petersen2015}. Nevertheless, we minimized this bias by following the guidelines for literature reviews~\citep{kitchenham2007, Petersen:2008, petersen2015, randolph2009guide}. Most notably, these include: defining the SLR protocol in advance, having two authors review the candidate papers independently, and discussing and resolving disagreements.

Even though our search criteria were broad, we selected only \numselectedpapers\ papers. This number suggests that the analysis of cybersecurity training artifacts is a narrow domain, and its community is not (yet) widely established. To support the development of this arising research area, we provide recommendations for further research in \Cref{sec:implications}. Nevertheless, the relatively small number of selected papers is typical for SLRs, which tend to have a narrow focus. In a comparison of ten literature reviews~\citep[p.~7]{Petersen:2008}, five of them inspected less than 30 papers.

\section{Implications of this Literature Survey}
\label{sec:implications}

Cybersecurity education research combined with EDM/LA has a substantial practical impact. It brings new insights into learning technologies and enables effective learning interventions. However, our SLR revealed that few studies fully exploit the potential of EDM/LA applied to student-generated data. One of the reasons could be that creating the training content itself is challenging, and few resources are left for other activities such as follow-up research.

To motivate further research, we formulate a research agenda in \Cref{subsec:future-work}. \Cref{subsec:tips} also provides recommendations for writing papers in the domain of EDM/LA to aid fellow cybersecurity education researchers. As a result, this SLR not only reviews the facts derived from existing literature but adds to the understanding of the area.

\subsection{Identified Research Gaps and Future Work Proposals}
\label{subsec:future-work}

We list several open problems not covered in the surveyed papers. To be specific, we phrase the problems as research questions and invite interested researchers to address them.
\begin{enumerate}
    \item The vast majority of the examined papers focused on offensive security and network security. An uncovered research question is: \textit{How can student data be leveraged to support other areas of cybersecurity education, such as secure programming, data security, or even human security?}
    \item Most reviewed papers performed a post-hoc analysis of student data. However, a real-time analysis would provide situational awareness and support classroom orchestration. Moreover, it would enable providing immediate automated feedback to students to improve their learning experience. Therefore, an interesting question is: \textit{Which information can be inferred from student data during the training to inform instructors and students about their progress?}
    \item A follow-up question to the previous one is: \textit{How to automatically adapt instruction based on the student data?} Similarly to intelligent tutoring systems, cybersecurity training environments can employ student data to personalize instruction according to the skill level of individual students. These technologies can reduce the barrier to participation of beginners.
    \item Since the median sample size was only 43 students, many statistical and machine learning methods are not applicable. However, the time span of data collection ranged from several hours to several days, during which each student generates in-depth data. So, it would be interesting to examine: \textit{Which automated methods for data analysis are suitable for a small number of students who interact with the training environment for a long time?}
    \item For researchers interested in writing literature survey papers, a relevant question is: \textit{How can cybersecurity be taught?} The review can examine the possible teaching methods, their effectiveness, advantages and disadvantages, and necessary infrastructure.
\end{enumerate}

Moreover, open problems in computing education research~\citep{denny2019} or general cybersecurity education~\citep{Svabensky2020} can also be applied to hands-on security training combined with EDM/LA.

Last but not least, cybersecurity education research requires an infrastructure for data collection. To support the research, developers of learning technologies can examine how to simplify the deployment of training environments that would enable seamless data collection and analysis.

Overall, there are many opportunities for fruitful future work. Research aimed at technologies that support learning has a great potential to improve the student experience. It can enable remote access to education at scale, provide rapid assessment and feedback, and reduce the burden placed on instructors.

\subsection{Recommendations for Publishing EDM/LA Research}
\label{subsec:tips}

Handbooks of EDM~\citep{handbook-edm2010} and LA~\citep{handbook-la2017} provide an excellent overview of general methods and research approaches, along with examples of studies. We wish to add more specific recommendations that we formulated while reviewing the \numselectedpapers\ papers. Therefore, we provide a list of six criteria that an EDM/LA paper in cybersecurity should address.
\begin{enumerate}
    \item Clearly describe the learning objectives of the cybersecurity educational intervention. Refer to a standardized curriculum, such as CSEC2017~\citep{cybered} or ACM/IEEE curricular guidelines~\citep{cc2020, sahami2013curricula}.
    \item Follow a thorough methodology for data collection.
    \begin{itemize}
        \item Explain the purpose of the data collection. Usually, this is bound to the studied research question. Are the data needed to assess students, help them learn, provide feedback to the training designers, or something else?
        \item Characterize the technical environment from which the data are collected. Is it a physical or virtual infrastructure? What would other researchers or instructors need to build it? Does it employ open-source components?
        \item List what data are collected, from which systems, and how they address the defined purpose.
        \item State the precise number of students (participants) from which the data were collected.
        \item Report the time span during which the data were collected.
        \item Explain the ethical measures. Which steps were taken to anonymize the data and preserve the privacy of participants? This usually involves informed consent or approval of the study by an institutional review board.
    \end{itemize}
    \item Select appropriate analysis methods suitable for the collected data and relevant for the defined purpose.
    \item Explicitly describe the contributions and practical applications of the research.
    \item When applying the research to practice, evaluate that the EDM/LA interventions helped improve some aspect of teaching or learning.
    \item If possible, publish the dataset, source code, configuration files, other relevant supplementary materials, and documentation on using or deploying them. This enables other researchers and instructors to replicate the setup and build upon the results.
\end{enumerate}

To increase the quality of cybersecurity education research papers, we also recommend the authors to study the methodology of computing education research~\citep{handbook-CER, Lishinski2016}. For the authors who want to apply inferential statistics to their data, we recommend the review of usage of statistics in computing education research~\citep{Sanders2019}. Finally, all authors who present analysis results visually can benefit from suggestions in~\citep{Simon2019}.

Regarding the publication of datasets, open-data initiatives such as Zenodo~\citep{zenodo} allow researchers to easily and permanently share their research data. When looking at specific communities of practice, activities such as The Graphics Replicability Stamp Initiative~\citep{grsi} encourage researchers in the field of computer graphics and visualizations to publish their research artifacts. A similar initiative would benefit the cybersecurity education community as well. It is challenging to deploy infrastructure for the training itself, not to mention enhancing it with data collection capabilities. Researchers who overcome this issue can publish their dataset and enable others to analyze it.

\section{Conclusions}
\label{sec:conclusion}

We surveyed publications that leverage student-generated data to support hands-on cybersecurity training. This interdisciplinary research area is still in its developing stages, so it is not yet well-understood. Our work helps to contextualize it and provides inspiration for researchers, practitioners, and educators.

We followed the best practices for systematic literature reviews, revealing diverse applications of EDM/LA methods on training artifacts. These range from understanding students' misconceptions, to developing tools for providing automated feedback, to evaluating assessment models for skill level prediction. As a result, EDM/LA research yields insights beneficial for students, instructors, and developers of interactive learning environments.

The emerging research in this area will also have a practical impact. Cybersecurity is a domain that needs educated experts -- millions of skilled workers now lack worldwide~\citep{isc2}. Examining new ways of employing student interaction data will enable a better understanding of teaching and learning processes and, ultimately, improve them. As a result, instructors will be better equipped to train cybersecurity specialists more efficiently. What is more, the surveyed methods are applicable in other domains of computing education, such as operating systems, networking, or programming.

Although our survey showed that student data from hands-on training have vast potential, researchers do not (yet) fully exploit it. We hope to support future research efforts in this area by providing the following contributions:
\begin{itemize}
    \item An organized inventory of papers along with the synthesis and classification of their approaches, results, and applications. This inventory evaluates the current trends and adds to the understanding of the area. It will aid both new researchers and those already in the field and inspire developers of cybersecurity learning technologies, instructors, and creators of training content.
    \item Identification of research trends as well as gaps, along with potential directions for future work to motivate further research.
    \item Practical recommendations for conducting cybersecurity education research.
\end{itemize}

As supplementary material~\citep{supplementary-materials} for the paper, we publish the raw dataset: BibTeX references exported from Mendeley reference manager~\citep{mendeley} that include citations of all candidate and selected papers. We also provide the processed dataset: a spreadsheet with the complete information about the \numselectedpapers\ selected papers.

\backmatter

\section*{Statements and Declarations}

\bmhead{Funding}
This research was supported by the ERDF project \textit{CyberSecurity, CyberCrime and Critical Information Infrastructures Center of Excellence} (No. CZ.02.1.01/0.0/0.0/16\_019/0000822).

\bmhead{Acknowledgments}
The authors thank cybersecurity experts from the CSIRT-MU team who helped select the keywords for the automated search. We also thank Radek Pelánek for his comments on the early stages of this article.

\bmhead{Competing interests}
The authors have no competing interests to declare.

\bmhead{Availability of data and materials}
The accompanying data and materials are published in a free, open-source repository on Zenodo~\citep{supplementary-materials}.

\bmhead{Authors' contributions}
\textit{Valdemar Švábenský}: Conceptualization, Methodology, Formal analysis, Investigation, Data Curation, Writing -- Original Draft, Visualization, Project administration. \textit{Jan Vykopal}: Conceptualization, Methodology, Investigation, Writing -- Review \& Editing, Data Curation. \textit{Pavel Čeleda}: Conceptualization, Writing -- Review \& Editing, Supervision, Funding acquisition. \textit{Lydia Kraus}: Investigation, Writing -- Review \& Editing

\bibliography{references}

\begin{thebibliography}{}
\providecommand{\doi}[1]{\url{https://doi.org/#1}}
\bibcommenthead

\bibitem [\protect \citeauthoryear {%
Abbott%
\ \protect \BOthers {.}}{%
Abbott%
\ \protect \BOthers {.}}{%
{\protect \APACyear {2015}}%
}]{%
Abbott2015log}
\APACinsertmetastar {%
Abbott2015log}%
\begin{APACrefauthors}%
Abbott, R.G.%
, McClain, J.%
, Anderson, B.%
, Nauer, K.%
, Silva, A.%
\BCBL {} Forsythe, C.%
\end{APACrefauthors}%
\unskip\
\newblock
\APACrefYearMonthDay{2015}{}{}.
\newblock
{\BBOQ}\APACrefatitle {{Log Analysis of Cyber Security Training Exercises}}
  {{Log Analysis of Cyber Security Training Exercises}}.{\BBCQ}
\newblock
\APACjournalVolNumPages{Procedia Manufacturing}{3}{}{5088--5094}.
\newblock
\begin{APACrefURL} {\url{https://doi.org/10.1016/j.promfg.2015.07.523}}
  \end{APACrefURL}
\newblock

\newblock

\PrintBackRefs{\CurrentBib}

\bibitem [\protect \citeauthoryear {%
Almansoori%
\ \protect \BOthers {.}}{%
Almansoori%
\ \protect \BOthers {.}}{%
{\protect \APACyear {2020}}%
}]{%
Almansoori2020}
\APACinsertmetastar {%
Almansoori2020}%
\begin{APACrefauthors}%
Almansoori, M.%
, Lam, J.%
, Fang, E.%
, Mulligan, K.%
, Soosai~Raj, A.G.%
\BCBL {} Chatterjee, R.%
\end{APACrefauthors}%
\unskip\
\newblock
\APACrefYearMonthDay{2020}{}{}.
\newblock
{\BBOQ}\APACrefatitle {{How Secure Are Our Computer Systems Courses?}} {{How
  Secure Are Our Computer Systems Courses?}}{\BBCQ}
\newblock
 \APACrefbtitle {{Proceedings of the 2020 ACM Conference on International
  Computing Education Research}} {{Proceedings of the 2020 ACM Conference on
  International Computing Education Research}}\ (\BPG~271–281).
\newblock
\APACaddressPublisher{New York, NY, USA}{Association for Computing Machinery}.
\newblock
\begin{APACrefURL} {\url{https://doi.org/10.1145/3372782.3406266}}
  \end{APACrefURL}
\PrintBackRefs{\CurrentBib}

\bibitem [\protect \citeauthoryear {%
Andreatos%
}{%
Andreatos%
}{%
{\protect \APACyear {2017}}%
}]{%
Andreatos2017}
\APACinsertmetastar {%
Andreatos2017}%
\begin{APACrefauthors}%
Andreatos, A.S.%
\end{APACrefauthors}%
\unskip\
\newblock
\APACrefYearMonthDay{2017}{}{}.
\newblock
{\BBOQ}\APACrefatitle {{Designing educational scenarios to teach network
  security}} {{Designing educational scenarios to teach network
  security}}.{\BBCQ}
\newblock
 \APACrefbtitle {{8th IEEE Global Engineering Education Conference, EDUCON
  2017}} {{8th IEEE Global Engineering Education Conference, EDUCON 2017}}\
  (\BPGS\ 1606--1610).
\newblock
\APACaddressPublisher{Washington, D.C., USA}{IEEE Computer Society}.
\newblock
\begin{APACrefURL} {\url{https://doi.org/10.1109/EDUCON.2017.7943063}}
  \end{APACrefURL}
\PrintBackRefs{\CurrentBib}

\bibitem [\protect \citeauthoryear {%
Andreolini%
, Colacino%
, Colajanni%
\BCBL {}\ \BBA {} Marchetti%
}{%
Andreolini%
\ \protect \BOthers {.}}{%
{\protect \APACyear {2019}}%
}]{%
Andreolini2019}
\APACinsertmetastar {%
Andreolini2019}%
\begin{APACrefauthors}%
Andreolini, M.%
, Colacino, V.G.%
, Colajanni, M.%
\BCBL {} Marchetti, M.%
\end{APACrefauthors}%
\unskip\
\newblock
\APACrefYearMonthDay{2019}{}{}.
\newblock
{\BBOQ}\APACrefatitle {{A Framework for the Evaluation of Trainee Performance
  in Cyber Range Exercises}} {{A Framework for the Evaluation of Trainee
  Performance in Cyber Range Exercises}}.{\BBCQ}
\newblock
\APACjournalVolNumPages{Mobile Networks and Applications}{25}{}{236--247}.
\newblock
\begin{APACrefURL} {\url{https://doi.org/10.1007/s11036-019-01442-0}}
  \end{APACrefURL}
\newblock

\newblock

\PrintBackRefs{\CurrentBib}

\bibitem [\protect \citeauthoryear {%
Burket%
, Chapman%
, Becker%
, Ganas%
\BCBL {}\ \BBA {} Brumley%
}{%
Burket%
\ \protect \BOthers {.}}{%
{\protect \APACyear {2015}}%
}]{%
Burket2015problemgeneration}
\APACinsertmetastar {%
Burket2015problemgeneration}%
\begin{APACrefauthors}%
Burket, J.%
, Chapman, P.%
, Becker, T.%
, Ganas, C.%
\BCBL {} Brumley, D.%
\end{APACrefauthors}%
\unskip\
\newblock
\APACrefYearMonthDay{2015}{}{}.
\newblock
{\BBOQ}\APACrefatitle {{Automatic Problem Generation for Capture-the-Flag
  Competitions}} {{Automatic Problem Generation for Capture-the-Flag
  Competitions}}.{\BBCQ}
\newblock
 \APACrefbtitle {{2015 {USENIX} Summit on Gaming, Games, and Gamification in
  Security Education (3GSE 15)}} {{2015 {USENIX} Summit on Gaming, Games, and
  Gamification in Security Education (3GSE 15)}}\ (\BPGS\ 1--8).
\newblock
\APACaddressPublisher{Berkeley, CA, USA}{{USENIX} Association}.
\newblock
\begin{APACrefURL}
  {\url{https://www.usenix.org/conference/3gse15/summit-program/presentation/burket}}
  \end{APACrefURL}
\PrintBackRefs{\CurrentBib}

\bibitem [\protect \citeauthoryear {%
Caliskan%
, Tatar%
, Bahsi%
, Ottis%
\BCBL {}\ \BBA {} Vaarandi%
}{%
Caliskan%
\ \protect \BOthers {.}}{%
{\protect \APACyear {2017}}%
}]{%
caliskan2017capability}
\APACinsertmetastar {%
caliskan2017capability}%
\begin{APACrefauthors}%
Caliskan, E.%
, Tatar, U.%
, Bahsi, H.%
, Ottis, R.%
\BCBL {} Vaarandi, R.%
\end{APACrefauthors}%
\unskip\
\newblock
\APACrefYearMonthDay{2017}{}{}.
\newblock
{\BBOQ}\APACrefatitle {Capability detection and evaluation metrics for cyber
  security lab exercises} {Capability detection and evaluation metrics for
  cyber security lab exercises}.{\BBCQ}
\newblock
 \APACrefbtitle {{Proceedings of the International Conference on Cyber Warfare
  and Security}} {{Proceedings of the International Conference on Cyber Warfare
  and Security}}\ (\BPGS\ 407--414).
\newblock
\APACaddressPublisher{Sonning Common, Reading, UK}{Academic Conferences and
  Publishing International}.
\PrintBackRefs{\CurrentBib}

\bibitem [\protect \citeauthoryear {%
{CC2020 Task Force}%
}{%
{CC2020 Task Force}%
}{%
{\protect \APACyear {2020}}%
}]{%
cc2020}
\APACinsertmetastar {%
cc2020}%
\begin{APACrefauthors}%
{CC2020 Task Force}%
\end{APACrefauthors}%
\unskip\
\newblock
\APACrefYear{2020}.
\newblock
\APACrefbtitle {{Computing Curricula 2020: Paradigms for Global Computing
  Education}} {{Computing Curricula 2020: Paradigms for Global Computing
  Education}}.
\newblock
\APACaddressPublisher{New York, NY, USA}{Association for Computing Machinery}.
\newblock
\begin{APACrefURL} {\url{https://doi.org/10.1145/3467967}} \end{APACrefURL}
\PrintBackRefs{\CurrentBib}

\bibitem [\protect \citeauthoryear {%
{CERN Data Centre \& Invenio}%
}{%
{CERN Data Centre \& Invenio}%
}{%
{\protect \APACyear {2022}}%
}]{%
zenodo}
\APACinsertmetastar {%
zenodo}%
\begin{APACrefauthors}%
{CERN Data Centre \& Invenio}%
\end{APACrefauthors}%
\unskip\
\newblock
\APACrefYearMonthDay{2022}{}{}.
\newblock
\APACrefbtitle {{Zenodo -- Research. Shared.}} {{Zenodo -- Research. Shared.}}
\newblock
\APAChowpublished {Retrieved April 6, 2022 from \url{https://zenodo.org}}.
\PrintBackRefs{\CurrentBib}

\bibitem [\protect \citeauthoryear {%
Chapman%
, Burket%
\BCBL {}\ \BBA {} Brumley%
}{%
Chapman%
\ \protect \BOthers {.}}{%
{\protect \APACyear {2014}}%
}]{%
chapman2014picoctf}
\APACinsertmetastar {%
chapman2014picoctf}%
\begin{APACrefauthors}%
Chapman, P.%
, Burket, J.%
\BCBL {} Brumley, D.%
\end{APACrefauthors}%
\unskip\
\newblock
\APACrefYearMonthDay{2014}{}{}.
\newblock
{\BBOQ}\APACrefatitle {{PicoCTF: A Game-Based Computer Security Competition for
  High School Students}} {{PicoCTF: A Game-Based Computer Security Competition
  for High School Students}}.{\BBCQ}
\newblock
 \APACrefbtitle {{2014 {USENIX} Summit on Gaming, Games, and Gamification in
  Security Education (3GSE 14)}} {{2014 {USENIX} Summit on Gaming, Games, and
  Gamification in Security Education (3GSE 14)}}\ (\BPGS\ 1--10).
\newblock
\APACaddressPublisher{Berkeley, CA, USA}{{USENIX} Association}.
\newblock
\begin{APACrefURL}
  {\url{https://www.usenix.org/conference/3gse14/summit-program/presentation/chapman}}
  \end{APACrefURL}
\PrintBackRefs{\CurrentBib}

\bibitem [\protect \citeauthoryear {%
Chothia%
, Holdcroft%
, Radu%
\BCBL {}\ \BBA {} Thomas%
}{%
Chothia%
\ \protect \BOthers {.}}{%
{\protect \APACyear {2017}}%
}]{%
Chothia2017jail}
\APACinsertmetastar {%
Chothia2017jail}%
\begin{APACrefauthors}%
Chothia, T.%
, Holdcroft, S.%
, Radu, A\BHBI I.%
\BCBL {} Thomas, R.J.%
\end{APACrefauthors}%
\unskip\
\newblock
\APACrefYearMonthDay{2017}{}{}.
\newblock
{\BBOQ}\APACrefatitle {{Jail, Hero or Drug Lord? Turning a Cyber Security
  Course Into an 11 Week Choose Your Own Adventure Story}} {{Jail, Hero or Drug
  Lord? Turning a Cyber Security Course Into an 11 Week Choose Your Own
  Adventure Story}}.{\BBCQ}
\newblock
 \APACrefbtitle {{2017 USENIX Workshop on Advances in Security Education (ASE
  17)}} {{2017 USENIX Workshop on Advances in Security Education (ASE 17)}}\
  (\BPGS\ 1--11).
\newblock
\APACaddressPublisher{Berkeley, CA, USA}{USENIX Association}.
\newblock
\begin{APACrefURL}
  {\url{https://www.usenix.org/conference/ase17/workshop-program/presentation/chothia}}
  \end{APACrefURL}
\PrintBackRefs{\CurrentBib}

\bibitem [\protect \citeauthoryear {%
Chothia%
\ \BBA {} Novakovic%
}{%
Chothia%
\ \BBA {} Novakovic%
}{%
{\protect \APACyear {2015}}%
}]{%
chothia2015offline}
\APACinsertmetastar {%
chothia2015offline}%
\begin{APACrefauthors}%
Chothia, T.%
\BCBT {}\ \BBA {} Novakovic, C.%
\end{APACrefauthors}%
\unskip\
\newblock
\APACrefYearMonthDay{2015}{}{}.
\newblock
{\BBOQ}\APACrefatitle {{An Offline Capture The Flag-Style Virtual Machine and
  an Assessment of Its Value for Cybersecurity Education}} {{An Offline Capture
  The Flag-Style Virtual Machine and an Assessment of Its Value for
  Cybersecurity Education}}.{\BBCQ}
\newblock
 \APACrefbtitle {{2015 {USENIX} Summit on Gaming, Games, and Gamification in
  Security Education (3GSE 15)}} {{2015 {USENIX} Summit on Gaming, Games, and
  Gamification in Security Education (3GSE 15)}}\ (\BPGS\ 1--8).
\newblock
\APACaddressPublisher{Berkeley, CA, USA}{{USENIX} Association}.
\newblock
\begin{APACrefURL}
  {\url{https://www.usenix.org/conference/3gse15/summit-program/presentation/chothia}}
  \end{APACrefURL}
\PrintBackRefs{\CurrentBib}

\bibitem [\protect \citeauthoryear {%
{Clarivate}%
}{%
{Clarivate}%
}{%
{\protect \APACyear {2022}}%
}]{%
jcr}
\APACinsertmetastar {%
jcr}%
\begin{APACrefauthors}%
{Clarivate}%
\end{APACrefauthors}%
\unskip\
\newblock
\APACrefYearMonthDay{2022}{}{}.
\newblock
\APACrefbtitle {{InCites Journal Citation Reports}.} {{InCites Journal Citation
  Reports}.}
\newblock
\APAChowpublished {Retrieved April 6, 2022 from
  \url{https://jcr.clarivate.com/jcr/home}}.
\PrintBackRefs{\CurrentBib}

\bibitem [\protect \citeauthoryear {%
{Computing Research and Education Association of Australasia}%
}{%
{Computing Research and Education Association of Australasia}%
}{%
{\protect \APACyear {2021}}%
}]{%
core}
\APACinsertmetastar {%
core}%
\begin{APACrefauthors}%
{Computing Research and Education Association of Australasia}%
\end{APACrefauthors}%
\unskip\
\newblock
\APACrefYearMonthDay{2021}{}{}.
\newblock
\APACrefbtitle {{CORE}.} {{CORE}.}
\newblock
\APAChowpublished {Retrieved April 6, 2022 from
  \url{http://portal.core.edu.au/conf-ranks/}}.
\PrintBackRefs{\CurrentBib}

\bibitem [\protect \citeauthoryear {%
{Cybersecurity \& Infrastructure Security Agency}%
}{%
{Cybersecurity \& Infrastructure Security Agency}%
}{%
{\protect \APACyear {2018}}%
}]{%
cyber-storm}
\APACinsertmetastar {%
cyber-storm}%
\begin{APACrefauthors}%
{Cybersecurity \& Infrastructure Security Agency}%
\end{APACrefauthors}%
\unskip\
\newblock
\APACrefYearMonthDay{2018}{}{}.
\newblock
\APACrefbtitle {{Cyber Storm: Securing Cyber Space}.} {{Cyber Storm: Securing
  Cyber Space}.}
\newblock
\APAChowpublished {Retrieved April 6, 2022 from
  \url{https://www.cisa.gov/cyber-storm-securing-cyber-space}}.
\PrintBackRefs{\CurrentBib}

\bibitem [\protect \citeauthoryear {%
{DEF CON}%
}{%
{DEF CON}%
}{%
{\protect \APACyear {2021}}%
}]{%
defconctf}
\APACinsertmetastar {%
defconctf}%
\begin{APACrefauthors}%
{DEF CON}%
\end{APACrefauthors}%
\unskip\
\newblock
\APACrefYearMonthDay{2021}{}{}.
\newblock
\APACrefbtitle {{CTF Archive}.} {{CTF Archive}.}
\newblock
\APAChowpublished {Retrieved April 6, 2022 from
  \url{https://www.defcon.org/html/links/dc-ctf.html}}.
\PrintBackRefs{\CurrentBib}

\bibitem [\protect \citeauthoryear {%
Deng%
, Lu%
, Chung%
, Huang%
\BCBL {}\ \BBA {} Zeng%
}{%
Deng%
\ \protect \BOthers {.}}{%
{\protect \APACyear {2018}}%
}]{%
Deng2018personalized}
\APACinsertmetastar {%
Deng2018personalized}%
\begin{APACrefauthors}%
Deng, Y.%
, Lu, D.%
, Chung, C\BHBI J.%
, Huang, D.%
\BCBL {} Zeng, Z.%
\end{APACrefauthors}%
\unskip\
\newblock
\APACrefYearMonthDay{2018}{}{}.
\newblock
{\BBOQ}\APACrefatitle {{Personalized Learning in a Virtual Hands-on Lab
  Platform for Computer Science Education}} {{Personalized Learning in a
  Virtual Hands-on Lab Platform for Computer Science Education}}.{\BBCQ}
\newblock
 \APACrefbtitle {{2018 IEEE Frontiers in Education Conference (FIE)}} {{2018
  IEEE Frontiers in Education Conference (FIE)}}\ (\BPGS\ 1--8).
\newblock
\APACaddressPublisher{New York, NY, USA}{IEEE}.
\newblock
\begin{APACrefURL} {\url{https://doi.org/10.1109/FIE.2018.8659291}}
  \end{APACrefURL}
\PrintBackRefs{\CurrentBib}

\bibitem [\protect \citeauthoryear {%
Denny%
, Becker%
, Craig%
, Wilson%
\BCBL {}\ \BBA {} Banaszkiewicz%
}{%
Denny%
\ \protect \BOthers {.}}{%
{\protect \APACyear {2019}}%
}]{%
denny2019}
\APACinsertmetastar {%
denny2019}%
\begin{APACrefauthors}%
Denny, P.%
, Becker, B.A.%
, Craig, M.%
, Wilson, G.%
\BCBL {} Banaszkiewicz, P.%
\end{APACrefauthors}%
\unskip\
\newblock
\APACrefYearMonthDay{2019}{}{}.
\newblock
{\BBOQ}\APACrefatitle {{Research This! Questions That Computing Educators Most
  Want Computing Education Researchers to Answer}} {{Research This! Questions
  That Computing Educators Most Want Computing Education Researchers to
  Answer}}.{\BBCQ}
\newblock
 \APACrefbtitle {{Proceedings of the 2019 ACM Conference on International
  Computing Education Research}} {{Proceedings of the 2019 ACM Conference on
  International Computing Education Research}}\ (\BPG~259–267).
\newblock
\APACaddressPublisher{New York, NY, USA}{Association for Computing Machinery}.
\newblock
\begin{APACrefURL} {\url{https://doi.org/10.1145/3291279.3339402}}
  \end{APACrefURL}
\PrintBackRefs{\CurrentBib}

\bibitem [\protect \citeauthoryear {%
{Elsevier}%
}{%
{Elsevier}%
}{%
{\protect \APACyear {2021}}%
}]{%
scopus}
\APACinsertmetastar {%
scopus}%
\begin{APACrefauthors}%
{Elsevier}%
\end{APACrefauthors}%
\unskip\
\newblock
\APACrefYearMonthDay{2021}{}{}.
\newblock
\APACrefbtitle {{Scopus}.} {{Scopus}.}
\newblock
\APAChowpublished {Retrieved April 6, 2022 from \url{https://www.scopus.com}}.
\PrintBackRefs{\CurrentBib}

\bibitem [\protect \citeauthoryear {%
Espinha~Gasiba%
, Lechner%
\BCBL {}\ \BBA {} Pinto-Albuquerque%
}{%
Espinha~Gasiba%
\ \protect \BOthers {.}}{%
{\protect \APACyear {2020}}%
}]{%
Gasiba2020}
\APACinsertmetastar {%
Gasiba2020}%
\begin{APACrefauthors}%
Espinha~Gasiba, T.%
, Lechner, U.%
\BCBL {} Pinto-Albuquerque, M.%
\end{APACrefauthors}%
\unskip\
\newblock
\APACrefYearMonthDay{2020}{11}{}.
\newblock
{\BBOQ}\APACrefatitle {{Cybersecurity Challenges in Industry: Measuring the
  Challenge Solve Time to Inform Future Challenges}} {{Cybersecurity Challenges
  in Industry: Measuring the Challenge Solve Time to Inform Future
  Challenges}}.{\BBCQ}
\newblock
\APACjournalVolNumPages{Information}{11}{11}{533}.
\newblock
\begin{APACrefURL} {\url{https://doi.org/10.3390/info11110533}}
  \end{APACrefURL}
\newblock

\newblock

\PrintBackRefs{\CurrentBib}

\bibitem [\protect \citeauthoryear {%
Estévez-Ayres%
, {Arias Fisteus}%
\BCBL {}\ \BBA {} Delgado-Kloos%
}{%
Estévez-Ayres%
\ \protect \BOthers {.}}{%
{\protect \APACyear {2017}}%
}]{%
Ayres2017}
\APACinsertmetastar {%
Ayres2017}%
\begin{APACrefauthors}%
Estévez-Ayres, I.%
, {Arias Fisteus}, J.%
\BCBL {} Delgado-Kloos, C.%
\end{APACrefauthors}%
\unskip\
\newblock
\APACrefYearMonthDay{2017}{}{}.
\newblock
{\BBOQ}\APACrefatitle {Lostrego: A distributed stream-based infrastructure for
  the real-time gathering and analysis of heterogeneous educational data}
  {Lostrego: A distributed stream-based infrastructure for the real-time
  gathering and analysis of heterogeneous educational data}.{\BBCQ}
\newblock
\APACjournalVolNumPages{{Journal of Network and Computer
  Applications}}{100}{}{56-68}.
\newblock
\begin{APACrefURL} {\url{https://doi.org/10.1016/j.jnca.2017.10.014}}
  \end{APACrefURL}
\newblock

\newblock

\PrintBackRefs{\CurrentBib}

\bibitem [\protect \citeauthoryear {%
Falah%
, Pan%
\BCBL {}\ \BBA {} Chen%
}{%
Falah%
\ \protect \BOthers {.}}{%
{\protect \APACyear {2019}}%
}]{%
Falah2019}
\APACinsertmetastar {%
Falah2019}%
\begin{APACrefauthors}%
Falah, A.%
, Pan, L.%
\BCBL {} Chen, F.%
\end{APACrefauthors}%
\unskip\
\newblock
\APACrefYearMonthDay{2019}{}{}.
\newblock
{\BBOQ}\APACrefatitle {{A Quantitative Approach to Design Special Purpose
  Systems to Measure Hacking Skills}} {{A Quantitative Approach to Design
  Special Purpose Systems to Measure Hacking Skills}}.{\BBCQ}
\newblock
 \APACrefbtitle {{2018 IEEE International Conference on Teaching, Assessment,
  and Learning for Engineering (TALE)}} {{2018 IEEE International Conference on
  Teaching, Assessment, and Learning for Engineering (TALE)}}\ (\BPGS\ 54--61).
\newblock
\APACaddressPublisher{New York, NY, USA}{IEEE}.
\newblock
\begin{APACrefURL} {\url{https://doi.org/10.1109/TALE.2018.8615431}}
  \end{APACrefURL}
\PrintBackRefs{\CurrentBib}

\bibitem [\protect \citeauthoryear {%
Fincher%
\ \BBA {} Robins%
}{%
Fincher%
\ \BBA {} Robins%
}{%
{\protect \APACyear {2019}}%
}]{%
handbook-CER}
\APACinsertmetastar {%
handbook-CER}%
\begin{APACrefauthors}%
Fincher, S.A.%
\BCBT {}\ \BBA {} Robins, A.V.%
\end{APACrefauthors}%
\ (\BEDS).
\unskip\
\newblock
\APACrefYear{2019}.
\newblock
\APACrefbtitle {{The Cambridge Handbook of Computing Education Research}} {{The
  Cambridge Handbook of Computing Education Research}}.
\newblock
\APACaddressPublisher{Cambridge, United Kingdom}{Cambridge University Press}.
\newblock
\begin{APACrefURL} {\url{https://doi.org/10.1017/9781108654555}}
  \end{APACrefURL}
\PrintBackRefs{\CurrentBib}

\bibitem [\protect \citeauthoryear {%
Google%
}{%
Google%
}{%
{\protect \APACyear {2021}}%
}]{%
googlectf}
\APACinsertmetastar {%
googlectf}%
\begin{APACrefauthors}%
Google%
\end{APACrefauthors}%
\unskip\
\newblock
\APACrefYearMonthDay{2021}{}{}.
\newblock
\APACrefbtitle {{Capture the Flag}.} {{Capture the Flag}.}
\newblock
\APAChowpublished {Retrieved April 6, 2022 from
  \url{https://capturetheflag.withgoogle.com}}.
\PrintBackRefs{\CurrentBib}

\bibitem [\protect \citeauthoryear {%
{Graham}%
\ \protect \BOthers {.}}{%
{Graham}%
\ \protect \BOthers {.}}{%
{\protect \APACyear {2020}}%
}]{%
Graham2020}
\APACinsertmetastar {%
Graham2020}%
\begin{APACrefauthors}%
{Graham}, K.%
, {Anderson}, J.%
, {Rife}, C.%
, {Heitmeyer}, B.%
, {R. Patel}, P.%
, {Nykl}, S.%
\BDBL {}{D. Merkle}, L.%
\end{APACrefauthors}%
\unskip\
\newblock
\APACrefYearMonthDay{2020}{}{}.
\newblock
{\BBOQ}\APACrefatitle {{Cyberspace Odyssey: A Competitive Team-Oriented Serious
  Game in Computer Networking}} {{Cyberspace Odyssey: A Competitive
  Team-Oriented Serious Game in Computer Networking}}.{\BBCQ}
\newblock
\APACjournalVolNumPages{{IEEE Transactions on Learning
  Technologies}}{13}{3}{502-515}.
\newblock
\begin{APACrefURL} {\url{https://doi.org/10.1109/TLT.2020.3008607}}
  \end{APACrefURL}
\newblock

\newblock

\PrintBackRefs{\CurrentBib}

\bibitem [\protect \citeauthoryear {%
Gran{\aa}sen%
\ \BBA {} Andersson%
}{%
Gran{\aa}sen%
\ \BBA {} Andersson%
}{%
{\protect \APACyear {2016}}%
}]{%
granaasen2016measuring}
\APACinsertmetastar {%
granaasen2016measuring}%
\begin{APACrefauthors}%
Gran{\aa}sen, M.%
\BCBT {}\ \BBA {} Andersson, D.%
\end{APACrefauthors}%
\unskip\
\newblock
\APACrefYearMonthDay{2016}{}{}.
\newblock
{\BBOQ}\APACrefatitle {Measuring team effectiveness in cyber-defense exercises:
  a cross-disciplinary case study} {Measuring team effectiveness in
  cyber-defense exercises: a cross-disciplinary case study}.{\BBCQ}
\newblock
\APACjournalVolNumPages{Cognition, Technology \& Work}{18}{1}{121--143}.
\newblock
\begin{APACrefURL} {\url{https://doi.org/10.1007/s10111-015-0350-2}}
  \end{APACrefURL}
\newblock

\newblock

\PrintBackRefs{\CurrentBib}

\bibitem [\protect \citeauthoryear {%
{Henshel}%
\ \protect \BOthers {.}}{%
{Henshel}%
\ \protect \BOthers {.}}{%
{\protect \APACyear {2016}}%
}]{%
Henshel2016predicting}
\APACinsertmetastar {%
Henshel2016predicting}%
\begin{APACrefauthors}%
{Henshel}, D.S.%
, {Deckard}, G.M.%
, {Lufkin}, B.%
, {Buchler}, N.%
, {Hoffman}, B.%
, {Rajivan}, P.%
\BCBL {} {Collman}, S.%
\end{APACrefauthors}%
\unskip\
\newblock
\APACrefYearMonthDay{2016}{}{}.
\newblock
{\BBOQ}\APACrefatitle {Predicting proficiency in cyber defense team exercises}
  {Predicting proficiency in cyber defense team exercises}.{\BBCQ}
\newblock
 \APACrefbtitle {{MILCOM 2016 -- IEEE Military Communications Conference}}
  {{MILCOM 2016 -- IEEE Military Communications Conference}}\ (\BPGS\
  776--781).
\newblock
\APACaddressPublisher{New York, NY, USA}{IEEE}.
\newblock
\begin{APACrefURL} {\url{https://doi.org/10.1109/MILCOM.2016.7795423}}
  \end{APACrefURL}
\PrintBackRefs{\CurrentBib}

\bibitem [\protect \citeauthoryear {%
Hundhausen%
, Olivares%
\BCBL {}\ \BBA {} Carter%
}{%
Hundhausen%
\ \protect \BOthers {.}}{%
{\protect \APACyear {2017}}%
}]{%
hundhausen2017}
\APACinsertmetastar {%
hundhausen2017}%
\begin{APACrefauthors}%
Hundhausen, C.%
, Olivares, D.%
\BCBL {} Carter, A.%
\end{APACrefauthors}%
\unskip\
\newblock
\APACrefYearMonthDay{2017}{08}{}.
\newblock
{\BBOQ}\APACrefatitle {{IDE-Based Learning Analytics for Computing Education: A
  Process Model, Critical Review, and Research Agenda}} {{IDE-Based Learning
  Analytics for Computing Education: A Process Model, Critical Review, and
  Research Agenda}}.{\BBCQ}
\newblock
\APACjournalVolNumPages{ACM Transactions on Computing
  Education}{17}{3}{11:1--11:26}.
\newblock
\begin{APACrefURL} {\url{https://doi.org/10.1145/3105759}} \end{APACrefURL}
\newblock

\newblock

\PrintBackRefs{\CurrentBib}

\bibitem [\protect \citeauthoryear {%
Ihantola%
\ \protect \BOthers {.}}{%
Ihantola%
\ \protect \BOthers {.}}{%
{\protect \APACyear {2015}}%
}]{%
ihantola2015}
\APACinsertmetastar {%
ihantola2015}%
\begin{APACrefauthors}%
Ihantola, P.%
, Vihavainen, A.%
, Ahadi, A.%
, Butler, M.%
, B\"{o}rstler, J.%
, Edwards, S.H.%
\BDBL {}Toll, D.%
\end{APACrefauthors}%
\unskip\
\newblock
\APACrefYearMonthDay{2015}{}{}.
\newblock
{\BBOQ}\APACrefatitle {{Educational Data Mining and Learning Analytics in
  Programming: Literature Review and Case Studies}} {{Educational Data Mining
  and Learning Analytics in Programming: Literature Review and Case
  Studies}}.{\BBCQ}
\newblock
 \APACrefbtitle {{Proceedings of the 2015 ITiCSE on Working Group Reports}}
  {{Proceedings of the 2015 ITiCSE on Working Group Reports}}\ (\BPGS\ 41--63).
\newblock
\APACaddressPublisher{New York, NY, USA}{ACM}.
\newblock
\begin{APACrefURL} {\url{https://doi.org/10.1145/2858796.2858798}}
  \end{APACrefURL}
\PrintBackRefs{\CurrentBib}

\bibitem [\protect \citeauthoryear {%
Imani%
\ \BBA {} Montazer%
}{%
Imani%
\ \BBA {} Montazer%
}{%
{\protect \APACyear {2019}}%
}]{%
Imani2019}
\APACinsertmetastar {%
Imani2019}%
\begin{APACrefauthors}%
Imani, M.%
\BCBT {}\ \BBA {} Montazer, G.A.%
\end{APACrefauthors}%
\unskip\
\newblock
\APACrefYearMonthDay{2019}{}{}.
\newblock
{\BBOQ}\APACrefatitle {A survey of emotion recognition methods with emphasis on
  E-Learning environments} {A survey of emotion recognition methods with
  emphasis on e-learning environments}.{\BBCQ}
\newblock
\APACjournalVolNumPages{Journal of Network and Computer
  Applications}{147}{}{102423}.
\newblock
\begin{APACrefURL} {\url{https://doi.org/10.1016/j.jnca.2019.102423}}
  \end{APACrefURL}
\newblock

\newblock

\PrintBackRefs{\CurrentBib}

\bibitem [\protect \citeauthoryear {%
{(ISC)\textsuperscript{2}}%
}{%
{(ISC)\textsuperscript{2}}%
}{%
{\protect \APACyear {2021}}%
}]{%
isc2}
\APACinsertmetastar {%
isc2}%
\begin{APACrefauthors}%
{(ISC)\textsuperscript{2}}%
\end{APACrefauthors}%
\unskip\
\newblock
\APACrefYearMonthDay{2021}{}{}.
\newblock
\APACrefbtitle {{Cybersecurity Workforce Study}} {{Cybersecurity Workforce
  Study}}\ \APACbVolEdTR{}{\BTR{}}.
\newblock
\begin{APACrefURL} {\url{https://www.isc2.org/Research/Workforce-Study}}
  \end{APACrefURL}
\PrintBackRefs{\CurrentBib}

\bibitem [\protect \citeauthoryear {%
{Joint Task Force on Computing Curricula, Association for Computing Machinery
  (ACM) and IEEE Computer Society}%
}{%
{Joint Task Force on Computing Curricula, Association for Computing Machinery
  (ACM) and IEEE Computer Society}%
}{%
{\protect \APACyear {2013}}%
}]{%
sahami2013curricula}
\APACinsertmetastar {%
sahami2013curricula}%
\begin{APACrefauthors}%
{Joint Task Force on Computing Curricula, Association for Computing Machinery
  (ACM) and IEEE Computer Society}%
\end{APACrefauthors}%
\unskip\
\newblock
\APACrefYear{2013}.
\newblock
\APACrefbtitle {{Computer Science Curricula 2013: Curriculum Guidelines for
  Undergraduate Degree Programs in Computer Science}} {{Computer Science
  Curricula 2013: Curriculum Guidelines for Undergraduate Degree Programs in
  Computer Science}}.
\newblock
\APACaddressPublisher{New York, NY, USA}{ACM}.
\newblock
\begin{APACrefURL} {\url{https://doi.org/10.1145/2534860}} \end{APACrefURL}
\PrintBackRefs{\CurrentBib}

\bibitem [\protect \citeauthoryear {%
{Joint Task Force on Cybersecurity Education}%
}{%
{Joint Task Force on Cybersecurity Education}%
}{%
{\protect \APACyear {2017}}%
}]{%
cybered}
\APACinsertmetastar {%
cybered}%
\begin{APACrefauthors}%
{Joint Task Force on Cybersecurity Education}%
\end{APACrefauthors}%
\unskip\
\newblock
\APACrefYearMonthDay{2017}{}{}.
\newblock
\APACrefbtitle {{Cybersecurity Curricular Guideline}.} {{Cybersecurity
  Curricular Guideline}.}
\newblock
\APAChowpublished {Retrieved April 6, 2022 from \url{https://cybered.acm.org}}.
\PrintBackRefs{\CurrentBib}

\bibitem [\protect \citeauthoryear {%
Kaneko%
\ \protect \BOthers {.}}{%
Kaneko%
\ \protect \BOthers {.}}{%
{\protect \APACyear {2020}}%
}]{%
Kaneko2020}
\APACinsertmetastar {%
Kaneko2020}%
\begin{APACrefauthors}%
Kaneko, K.%
, Igarashi, T.%
, Kayama, K.%
, Takeuchi, T.%
, Suzuki, T.%
, Kawase, A.%
\BDBL {}Okamura, K.%
\end{APACrefauthors}%
\unskip\
\newblock
\APACrefYearMonthDay{2020}{}{}.
\newblock
{\BBOQ}\APACrefatitle {{Learning Analytics with Multi-faced Data for
  Cybersecurity Education}} {{Learning Analytics with Multi-faced Data for
  Cybersecurity Education}}.{\BBCQ}
\newblock
 \APACrefbtitle {{9th International Congress on Advanced Applied Informatics
  (IIAI-AAI)}} {{9th International Congress on Advanced Applied Informatics
  (IIAI-AAI)}}\ (\BPGS\ 244--249).
\newblock
\begin{APACrefURL} {\url{https://doi.org/10.1109/IIAI-AAI50415.2020.00055}}
  \end{APACrefURL}
\PrintBackRefs{\CurrentBib}

\bibitem [\protect \citeauthoryear {%
Kasurinen%
\ \BBA {} Knutas%
}{%
Kasurinen%
\ \BBA {} Knutas%
}{%
{\protect \APACyear {2018}}%
}]{%
Kasurinen2018}
\APACinsertmetastar {%
Kasurinen2018}%
\begin{APACrefauthors}%
Kasurinen, J.%
\BCBT {}\ \BBA {} Knutas, A.%
\end{APACrefauthors}%
\unskip\
\newblock
\APACrefYearMonthDay{2018}{}{}.
\newblock
{\BBOQ}\APACrefatitle {Publication trends in gamification: A systematic mapping
  study} {Publication trends in gamification: A systematic mapping
  study}.{\BBCQ}
\newblock
\APACjournalVolNumPages{Elsevier Computer Science Review}{27}{}{33--44}.
\newblock
\begin{APACrefURL} {\url{https://doi.org/10.1016/j.cosrev.2017.10.003}}
  \end{APACrefURL}
\newblock

\newblock

\PrintBackRefs{\CurrentBib}

\bibitem [\protect \citeauthoryear {%
Khando%
, Gao%
, Islam%
\BCBL {}\ \BBA {} Salman%
}{%
Khando%
\ \protect \BOthers {.}}{%
{\protect \APACyear {2021}}%
}]{%
Khando2021}
\APACinsertmetastar {%
Khando2021}%
\begin{APACrefauthors}%
Khando, K.%
, Gao, S.%
, Islam, S.M.%
\BCBL {} Salman, A.%
\end{APACrefauthors}%
\unskip\
\newblock
\APACrefYearMonthDay{2021}{}{}.
\newblock
{\BBOQ}\APACrefatitle {Enhancing employees information security awareness in
  private and public organisations: A systematic literature review} {Enhancing
  employees information security awareness in private and public organisations:
  A systematic literature review}.{\BBCQ}
\newblock
\APACjournalVolNumPages{Computers \& Security}{106}{}{102267}.
\newblock
\begin{APACrefURL} {\url{https://doi.org/10.1016/j.cose.2021.102267}}
  \end{APACrefURL}
\newblock

\newblock

\PrintBackRefs{\CurrentBib}

\bibitem [\protect \citeauthoryear {%
Kitchenham%
\ \BBA {} Charters%
}{%
Kitchenham%
\ \BBA {} Charters%
}{%
{\protect \APACyear {2007}}%
}]{%
kitchenham2007}
\APACinsertmetastar {%
kitchenham2007}%
\begin{APACrefauthors}%
Kitchenham, B.%
\BCBT {}\ \BBA {} Charters, S.%
\end{APACrefauthors}%
\unskip\
\newblock
\APACrefYearMonthDay{2007}{}{}.
\newblock
\APACrefbtitle {Guidelines for performing Systematic Literature Reviews in
  Software Engineering} {Guidelines for performing systematic literature
  reviews in software engineering}\ \APACbVolEdTR{}{\BTR{}}.
\newblock
\APACaddressInstitution{}{EBSE}.
\PrintBackRefs{\CurrentBib}

\bibitem [\protect \citeauthoryear {%
{Knobbout}%
\ \BBA {} {Van Der Stappen}%
}{%
{Knobbout}%
\ \BBA {} {Van Der Stappen}%
}{%
{\protect \APACyear {2020}}%
}]{%
Knobbout2020}
\APACinsertmetastar {%
Knobbout2020}%
\begin{APACrefauthors}%
{Knobbout}, J.%
\BCBT {}\ \BBA {} {Van Der Stappen}, E.%
\end{APACrefauthors}%
\unskip\
\newblock
\APACrefYearMonthDay{2020}{}{}.
\newblock
{\BBOQ}\APACrefatitle {{Where is the Learning in Learning Analytics? A
  Systematic Literature Review on the Operationalization of Learning-Related
  Constructs in the Evaluation of Learning Analytics Interventions}} {{Where is
  the Learning in Learning Analytics? A Systematic Literature Review on the
  Operationalization of Learning-Related Constructs in the Evaluation of
  Learning Analytics Interventions}}.{\BBCQ}
\newblock
\APACjournalVolNumPages{IEEE Transactions on Learning
  Technologies}{13}{3}{631--645}.
\newblock
\begin{APACrefURL} {\url{https://doi.org/10.1109/TLT.2020.2999970}}
  \end{APACrefURL}
\newblock

\newblock

\PrintBackRefs{\CurrentBib}

\bibitem [\protect \citeauthoryear {%
Kokkonen%
\ \BBA {} Puuska%
}{%
Kokkonen%
\ \BBA {} Puuska%
}{%
{\protect \APACyear {2018}}%
}]{%
Kokkonen2018}
\APACinsertmetastar {%
Kokkonen2018}%
\begin{APACrefauthors}%
Kokkonen, T.%
\BCBT {}\ \BBA {} Puuska, S.%
\end{APACrefauthors}%
\unskip\
\newblock
\APACrefYearMonthDay{2018}{}{}.
\newblock
{\BBOQ}\APACrefatitle {{Blue Team Communication and Reporting for Enhancing
  Situational Awareness from White Team Perspective in Cyber Security
  Exercises}} {{Blue Team Communication and Reporting for Enhancing Situational
  Awareness from White Team Perspective in Cyber Security Exercises}}.{\BBCQ}
\newblock
 \APACrefbtitle {{18th International Conference on Next Generation Teletraffic
  and Wired/Wireless Advanced Networks and Systems, NEW2AN 2018 and 11th
  Conference on Internet of Things and Smart Spaces, ruSMART 2018}} {{18th
  International Conference on Next Generation Teletraffic and Wired/Wireless
  Advanced Networks and Systems, NEW2AN 2018 and 11th Conference on Internet of
  Things and Smart Spaces, ruSMART 2018}}\ (\BVOL\ 11118 LNCS, \BPGS\
  277--288).
\newblock
\APACaddressPublisher{Vienna, Austria}{Springer}.
\newblock
\begin{APACrefURL} {\url{https://doi.org/10.1007/978-3-030-01168-0_26}}
  \end{APACrefURL}
\PrintBackRefs{\CurrentBib}

\bibitem [\protect \citeauthoryear {%
Kont%
, Pihelgas%
, Maennel%
, Blumbergs%
\BCBL {}\ \BBA {} Lepik%
}{%
Kont%
\ \protect \BOthers {.}}{%
{\protect \APACyear {2017}}%
}]{%
Kont2017}
\APACinsertmetastar {%
Kont2017}%
\begin{APACrefauthors}%
Kont, M.%
, Pihelgas, M.%
, Maennel, K.%
, Blumbergs, B.%
\BCBL {} Lepik, T.%
\end{APACrefauthors}%
\unskip\
\newblock
\APACrefYearMonthDay{2017}{}{}.
\newblock
{\BBOQ}\APACrefatitle {{Frankenstack: Toward real-time Red Team feedback}}
  {{Frankenstack: Toward real-time Red Team feedback}}.{\BBCQ}
\newblock
 \APACrefbtitle {{2017 IEEE Military Communications Conference, MILCOM 2017}}
  {{2017 IEEE Military Communications Conference, MILCOM 2017}}\ (\BPGS\
  400--405).
\newblock
\APACaddressPublisher{New York, NY, USA}{IEEE}.
\newblock
\begin{APACrefURL} {\url{https://doi.org/10.1109/MILCOM.2017.8170852}}
  \end{APACrefURL}
\PrintBackRefs{\CurrentBib}

\bibitem [\protect \citeauthoryear {%
Krippendorff%
}{%
Krippendorff%
}{%
{\protect \APACyear {2004}}%
}]{%
krippendorff2004reliability}
\APACinsertmetastar {%
krippendorff2004reliability}%
\begin{APACrefauthors}%
Krippendorff, K.%
\end{APACrefauthors}%
\unskip\
\newblock
\APACrefYearMonthDay{2004}{}{}.
\newblock
{\BBOQ}\APACrefatitle {Reliability in content analysis: Some common
  misconceptions and recommendations} {Reliability in content analysis: Some
  common misconceptions and recommendations}.{\BBCQ}
\newblock
\APACjournalVolNumPages{Human communication research}{30}{3}{411--433}.
\newblock
\begin{APACrefURL} {\url{https://doi.org/10.1111/j.1468-2958.2004.tb00738.x}}
  \end{APACrefURL}
\newblock

\newblock

\PrintBackRefs{\CurrentBib}

\bibitem [\protect \citeauthoryear {%
Kucek%
\ \BBA {} Leitner%
}{%
Kucek%
\ \BBA {} Leitner%
}{%
{\protect \APACyear {2020}}%
}]{%
Kucek2020}
\APACinsertmetastar {%
Kucek2020}%
\begin{APACrefauthors}%
Kucek, S.%
\BCBT {}\ \BBA {} Leitner, M.%
\end{APACrefauthors}%
\unskip\
\newblock
\APACrefYearMonthDay{2020}{}{}.
\newblock
{\BBOQ}\APACrefatitle {{An Empirical Survey of Functions and Configurations of
  Open-Source Capture the Flag (CTF) Environments}} {{An Empirical Survey of
  Functions and Configurations of Open-Source Capture the Flag (CTF)
  Environments}}.{\BBCQ}
\newblock
\APACjournalVolNumPages{Journal of Network and Computer Applications}{151}{}{}.
\newblock
\begin{APACrefURL} {\url{https://doi.org/10.1016/j.jnca.2019.102470}}
  \end{APACrefURL}
\newblock

\newblock

\PrintBackRefs{\CurrentBib}

\bibitem [\protect \citeauthoryear {%
Labuschagne%
\ \BBA {} Grobler%
}{%
Labuschagne%
\ \BBA {} Grobler%
}{%
{\protect \APACyear {2017}}%
}]{%
Labuschagne2017}
\APACinsertmetastar {%
Labuschagne2017}%
\begin{APACrefauthors}%
Labuschagne, W.A.%
\BCBT {}\ \BBA {} Grobler, M.%
\end{APACrefauthors}%
\unskip\
\newblock
\APACrefYearMonthDay{2017}{}{}.
\newblock
{\BBOQ}\APACrefatitle {{Developing a capability to classify technical skill
  levels within a cyber range}} {{Developing a capability to classify technical
  skill levels within a cyber range}}.{\BBCQ}
\newblock
 \APACrefbtitle {{16th European Conference on Cyber Warfare and Security, ECCWS
  2017}} {{16th European Conference on Cyber Warfare and Security, ECCWS
  2017}}\ (\BPGS\ 224--234).
\newblock
\APACaddressPublisher{Red Hook, NY, USA}{Curran Associates Inc.}
\newblock
\begin{APACrefURL} {\url{https://www.proquest.com/docview/1966803837}}
  \end{APACrefURL}
\PrintBackRefs{\CurrentBib}

\bibitem [\protect \citeauthoryear {%
Lang%
, Siemens%
, Wise%
\BCBL {}\ \BBA {} Gašević%
}{%
Lang%
\ \protect \BOthers {.}}{%
{\protect \APACyear {2017}}%
}]{%
handbook-la2017}
\APACinsertmetastar {%
handbook-la2017}%
\begin{APACrefauthors}%
Lang, C.%
, Siemens, G.%
, Wise, A.%
\BCBL {}\ \BBA {} Gašević, D.%
\end{APACrefauthors}%
\ (\BEDS).
\unskip\
\newblock
\APACrefYear{2017}.
\newblock
\APACrefbtitle {Handbook of Learning Analytics} {Handbook of learning
  analytics}\ (\PrintOrdinal{1st}\ \BEd).
\newblock
\APACaddressPublisher{}{Society for Learning Analytics Research (SoLAR)}.
\newblock
\begin{APACrefURL} {\url{https://doi.org/10.18608/hla17}} \end{APACrefURL}
\PrintBackRefs{\CurrentBib}

\bibitem [\protect \citeauthoryear {%
Li{\~n}{\'a}n%
\ \BBA {} P{\'e}rez%
}{%
Li{\~n}{\'a}n%
\ \BBA {} P{\'e}rez%
}{%
{\protect \APACyear {2015}}%
}]{%
linan2015}
\APACinsertmetastar {%
linan2015}%
\begin{APACrefauthors}%
Li{\~n}{\'a}n, L.C.%
\BCBT {}\ \BBA {} P{\'e}rez, {\'A}.A.J.%
\end{APACrefauthors}%
\unskip\
\newblock
\APACrefYearMonthDay{2015}{}{}.
\newblock
{\BBOQ}\APACrefatitle {Educational Data Mining and Learning Analytics:
  differences, similarities, and time evolution} {Educational data mining and
  learning analytics: differences, similarities, and time evolution}.{\BBCQ}
\newblock
\APACjournalVolNumPages{International Journal of Educational Technology in
  Higher Education}{12}{3}{98--112}.
\newblock
\begin{APACrefURL} {\url{https://doi.org/10.7238/rusc.v12i3.2515}}
  \end{APACrefURL}
\newblock

\newblock

\PrintBackRefs{\CurrentBib}

\bibitem [\protect \citeauthoryear {%
Lishinski%
, Good%
, Sands%
\BCBL {}\ \BBA {} Yadav%
}{%
Lishinski%
\ \protect \BOthers {.}}{%
{\protect \APACyear {2016}}%
}]{%
Lishinski2016}
\APACinsertmetastar {%
Lishinski2016}%
\begin{APACrefauthors}%
Lishinski, A.%
, Good, J.%
, Sands, P.%
\BCBL {} Yadav, A.%
\end{APACrefauthors}%
\unskip\
\newblock
\APACrefYearMonthDay{2016}{}{}.
\newblock
{\BBOQ}\APACrefatitle {{Methodological Rigor and Theoretical Foundations of CS
  Education Research}} {{Methodological Rigor and Theoretical Foundations of CS
  Education Research}}.{\BBCQ}
\newblock
 \APACrefbtitle {{Proceedings of the 2016 ACM Conference on International
  Computing Education Research}} {{Proceedings of the 2016 ACM Conference on
  International Computing Education Research}}\ (\BPG~161–169).
\newblock
\APACaddressPublisher{New York, NY, USA}{Association for Computing Machinery}.
\newblock
\begin{APACrefURL} {\url{https://doi.org/10.1145/2960310.2960328}}
  \end{APACrefURL}
\PrintBackRefs{\CurrentBib}

\bibitem [\protect \citeauthoryear {%
Luxton-Reilly%
\ \protect \BOthers {.}}{%
Luxton-Reilly%
\ \protect \BOthers {.}}{%
{\protect \APACyear {2018}}%
}]{%
Luxton-Reilly:2018}
\APACinsertmetastar {%
Luxton-Reilly:2018}%
\begin{APACrefauthors}%
Luxton-Reilly, A.%
, Simon%
, Albluwi, I.%
, Becker, B.A.%
, Giannakos, M.%
, Kumar, A.N.%
\BDBL {}Szabo, C.%
\end{APACrefauthors}%
\unskip\
\newblock
\APACrefYearMonthDay{2018}{}{}.
\newblock
{\BBOQ}\APACrefatitle {{Introductory Programming: A Systematic Literature
  Review}} {{Introductory Programming: A Systematic Literature Review}}.{\BBCQ}
\newblock
 \APACrefbtitle {{Proceedings Companion of the 23rd Annual ACM Conference on
  Innovation and Technology in Computer Science Education}} {{Proceedings
  Companion of the 23rd Annual ACM Conference on Innovation and Technology in
  Computer Science Education}}\ (\BPGS\ 55--106).
\newblock
\APACaddressPublisher{New York, NY, USA}{ACM}.
\newblock
\begin{APACrefURL} {\url{https://doi.org/10.1145/3293881.3295779}}
  \end{APACrefURL}
\PrintBackRefs{\CurrentBib}

\bibitem [\protect \citeauthoryear {%
{Maennel}%
}{%
{Maennel}%
}{%
{\protect \APACyear {2020}}%
}]{%
Maennel2020}
\APACinsertmetastar {%
Maennel2020}%
\begin{APACrefauthors}%
{Maennel}, K.%
\end{APACrefauthors}%
\unskip\
\newblock
\APACrefYearMonthDay{2020}{}{}.
\newblock
{\BBOQ}\APACrefatitle {{Learning Analytics Perspective: Evidencing Learning
  from Digital Datasets in Cybersecurity Exercises}} {{Learning Analytics
  Perspective: Evidencing Learning from Digital Datasets in Cybersecurity
  Exercises}}.{\BBCQ}
\newblock
 \APACrefbtitle {{2020 IEEE European Symposium on Security and Privacy
  Workshops (EuroSPW)}} {{2020 IEEE European Symposium on Security and Privacy
  Workshops (EuroSPW)}}\ (\BPGS\ 27--36).
\newblock
\begin{APACrefURL} {\url{https://doi.org/10.1109/EuroSPW51379.2020.00013}}
  \end{APACrefURL}
\PrintBackRefs{\CurrentBib}

\bibitem [\protect \citeauthoryear {%
Maennel%
, Mäses%
, Sütterlin%
, Ernits%
\BCBL {}\ \BBA {} Maennel%
}{%
Maennel%
\ \protect \BOthers {.}}{%
{\protect \APACyear {2019}}%
}]{%
Maennel2019}
\APACinsertmetastar {%
Maennel2019}%
\begin{APACrefauthors}%
Maennel, K.%
, Mäses, S.%
, Sütterlin, S.%
, Ernits, M.%
\BCBL {} Maennel, O.%
\end{APACrefauthors}%
\unskip\
\newblock
\APACrefYearMonthDay{2019}{}{}.
\newblock
{\BBOQ}\APACrefatitle {Using Technical Cybersecurity Exercises in University
  Admissions and Skill Evaluation} {Using technical cybersecurity exercises in
  university admissions and skill evaluation}.{\BBCQ}
\newblock
\APACjournalVolNumPages{IFAC-PapersOnLine}{52}{19}{169-174}.
\newblock
\begin{APACrefURL} {\url{https://doi.org/10.1016/j.ifacol.2019.12.169}}
  \end{APACrefURL}
\newblock
\APACrefnote{{14th IFAC Symposium on Analysis, Design, and Evaluation of Human
  Machine Systems (HMS 2019)}}
\newblock

\newblock

\PrintBackRefs{\CurrentBib}

\bibitem [\protect \citeauthoryear {%
Maennel%
, Ottis%
\BCBL {}\ \BBA {} Maennel%
}{%
Maennel%
\ \protect \BOthers {.}}{%
{\protect \APACyear {2017}}%
}]{%
Maennel2017}
\APACinsertmetastar {%
Maennel2017}%
\begin{APACrefauthors}%
Maennel, K.%
, Ottis, R.%
\BCBL {} Maennel, O.%
\end{APACrefauthors}%
\unskip\
\newblock
\APACrefYearMonthDay{2017}{}{}.
\newblock
{\BBOQ}\APACrefatitle {{Improving and measuring learning effectiveness at cyber
  defense exercises}} {{Improving and measuring learning effectiveness at cyber
  defense exercises}}.{\BBCQ}
\newblock
 \APACrefbtitle {{22nd Nordic Conference on Secure IT Systems, NordSec 2017}}
  {{22nd Nordic Conference on Secure IT Systems, NordSec 2017}}\ (\BPGS\
  123--138).
\newblock
\APACaddressPublisher{Vienna, Austria}{Springer}.
\newblock
\begin{APACrefURL} {\url{https://doi.org/10.1007/978-3-319-70290-2_8}}
  \end{APACrefURL}
\PrintBackRefs{\CurrentBib}

\bibitem [\protect \citeauthoryear {%
Malmi%
\ \protect \BOthers {.}}{%
Malmi%
\ \protect \BOthers {.}}{%
{\protect \APACyear {2010}}%
}]{%
Malmi2010}
\APACinsertmetastar {%
Malmi2010}%
\begin{APACrefauthors}%
Malmi, L.%
, Sheard, J.%
, Simon%
, Bednarik, R.%
, Helminen, J.%
, Korhonen, A.%
\BDBL {}Taherkhani, A.%
\end{APACrefauthors}%
\unskip\
\newblock
\APACrefYearMonthDay{2010}{}{}.
\newblock
{\BBOQ}\APACrefatitle {{Characterizing Research in Computing Education: A
  Preliminary Analysis of the Literature}} {{Characterizing Research in
  Computing Education: A Preliminary Analysis of the Literature}}.{\BBCQ}
\newblock
 \APACrefbtitle {{Proceedings of the Sixth International Workshop on Computing
  Education Research}} {{Proceedings of the Sixth International Workshop on
  Computing Education Research}}\ (\BPGS\ 3--12).
\newblock
\APACaddressPublisher{New York, NY, USA}{Association for Computing Machinery}.
\newblock
\begin{APACrefURL} {\url{https://doi.org/10.1145/1839594.1839597}}
  \end{APACrefURL}
\PrintBackRefs{\CurrentBib}

\bibitem [\protect \citeauthoryear {%
{Mangaroska}%
\ \BBA {} {Giannakos}%
}{%
{Mangaroska}%
\ \BBA {} {Giannakos}%
}{%
{\protect \APACyear {2019}}%
}]{%
Mangaroska2019}
\APACinsertmetastar {%
Mangaroska2019}%
\begin{APACrefauthors}%
{Mangaroska}, K.%
\BCBT {}\ \BBA {} {Giannakos}, M.%
\end{APACrefauthors}%
\unskip\
\newblock
\APACrefYearMonthDay{2019}{}{}.
\newblock
{\BBOQ}\APACrefatitle {{Learning Analytics for Learning Design: A Systematic
  Literature Review of Analytics-Driven Design to Enhance Learning}} {{Learning
  Analytics for Learning Design: A Systematic Literature Review of
  Analytics-Driven Design to Enhance Learning}}.{\BBCQ}
\newblock
\APACjournalVolNumPages{IEEE Transactions on Learning
  Technologies}{12}{4}{516--534}.
\newblock
\begin{APACrefURL} {\url{https://doi.org/10.1109/TLT.2018.2868673}}
  \end{APACrefURL}
\newblock

\newblock

\PrintBackRefs{\CurrentBib}

\bibitem [\protect \citeauthoryear {%
Margulieux%
, Ketenci%
\BCBL {}\ \BBA {} Decker%
}{%
Margulieux%
\ \protect \BOthers {.}}{%
{\protect \APACyear {2019}}%
}]{%
Margulieux2019}
\APACinsertmetastar {%
Margulieux2019}%
\begin{APACrefauthors}%
Margulieux, L.%
, Ketenci, T.A.%
\BCBL {} Decker, A.%
\end{APACrefauthors}%
\unskip\
\newblock
\APACrefYearMonthDay{2019}{}{}.
\newblock
{\BBOQ}\APACrefatitle {Review of measurements used in computing education
  research and suggestions for increasing standardization} {Review of
  measurements used in computing education research and suggestions for
  increasing standardization}.{\BBCQ}
\newblock
\APACjournalVolNumPages{Computer Science Education}{29}{1}{49--78}.
\newblock
\begin{APACrefURL} {\url{https://doi.org/10.1080/08993408.2018.1562145}}
  \end{APACrefURL}
\newblock

\newblock

\PrintBackRefs{\CurrentBib}

\bibitem [\protect \citeauthoryear {%
{Matcha}%
, {Uzir}%
, {Gašević}%
\BCBL {}\ \BBA {} {Pardo}%
}{%
{Matcha}%
\ \protect \BOthers {.}}{%
{\protect \APACyear {2020}}%
}]{%
Matcha2020}
\APACinsertmetastar {%
Matcha2020}%
\begin{APACrefauthors}%
{Matcha}, W.%
, {Uzir}, N.A.%
, {Gašević}, D.%
\BCBL {} {Pardo}, A.%
\end{APACrefauthors}%
\unskip\
\newblock
\APACrefYearMonthDay{2020}{}{}.
\newblock
{\BBOQ}\APACrefatitle {{A Systematic Review of Empirical Studies on Learning
  Analytics Dashboards: A Self-Regulated Learning Perspective}} {{A Systematic
  Review of Empirical Studies on Learning Analytics Dashboards: A
  Self-Regulated Learning Perspective}}.{\BBCQ}
\newblock
\APACjournalVolNumPages{IEEE Transactions on Learning
  Technologies}{13}{2}{226--245}.
\newblock
\begin{APACrefURL} {\url{https://doi.org/10.1109/TLT.2019.2916802}}
  \end{APACrefURL}
\newblock

\newblock

\PrintBackRefs{\CurrentBib}

\bibitem [\protect \citeauthoryear {%
{Mendeley}%
}{%
{Mendeley}%
}{%
{\protect \APACyear {2021}}%
}]{%
mendeley}
\APACinsertmetastar {%
mendeley}%
\begin{APACrefauthors}%
{Mendeley}%
\end{APACrefauthors}%
\unskip\
\newblock
\APACrefYearMonthDay{2021}{}{}.
\newblock
\APACrefbtitle {{Reference Manager}.} {{Reference Manager}.}
\newblock
\APAChowpublished {Retrieved April 6, 2022 from
  \url{https://www.mendeley.com/reference-management/reference-manager}}.
\PrintBackRefs{\CurrentBib}

\bibitem [\protect \citeauthoryear {%
Moher%
, Liberati%
, Tetzlaff%
, Altman%
\BCBL {}\ \BBA {} Group%
}{%
Moher%
\ \protect \BOthers {.}}{%
{\protect \APACyear {2009}}%
}]{%
prisma2009}
\APACinsertmetastar {%
prisma2009}%
\begin{APACrefauthors}%
Moher, D.%
, Liberati, A.%
, Tetzlaff, J.%
, Altman, D.G.%
\BCBL {} Group, T.P.%
\end{APACrefauthors}%
\unskip\
\newblock
\APACrefYearMonthDay{2009}{07}{}.
\newblock
{\BBOQ}\APACrefatitle {{Preferred Reporting Items for Systematic Reviews and
  Meta-Analyses: The PRISMA Statement}} {{Preferred Reporting Items for
  Systematic Reviews and Meta-Analyses: The PRISMA Statement}}.{\BBCQ}
\newblock
\APACjournalVolNumPages{PLOS Medicine}{6}{7}{1-6}.
\newblock
\begin{APACrefURL} {\url{https://doi.org/10.1371/journal.pmed.1000097}}
  \end{APACrefURL}
\newblock

\newblock

\PrintBackRefs{\CurrentBib}

\bibitem [\protect \citeauthoryear {%
{Nadeem}%
, {Allen}%
\BCBL {}\ \BBA {} {Williams}%
}{%
{Nadeem}%
\ \protect \BOthers {.}}{%
{\protect \APACyear {2015}}%
}]{%
Nadeem2015method}
\APACinsertmetastar {%
Nadeem2015method}%
\begin{APACrefauthors}%
{Nadeem}, M.%
, {Allen}, E.B.%
\BCBL {} {Williams}, B.J.%
\end{APACrefauthors}%
\unskip\
\newblock
\APACrefYearMonthDay{2015}{}{}.
\newblock
{\BBOQ}\APACrefatitle {{A Method for Recommending Computer-Security Training
  for Software Developers: Leveraging the Power of Static Analysis Techniques
  and Vulnerability Repositories}} {{A Method for Recommending
  Computer-Security Training for Software Developers: Leveraging the Power of
  Static Analysis Techniques and Vulnerability Repositories}}.{\BBCQ}
\newblock
 \APACrefbtitle {{12th International Conference on Information Technology --
  New Generations}} {{12th International Conference on Information Technology
  -- New Generations}}\ (\BPGS\ 534--539).
\newblock
\APACaddressPublisher{New York, NY, USA}{IEEE}.
\newblock
\begin{APACrefURL} {\url{https://doi.org/10.1109/ITNG.2015.90}}
  \end{APACrefURL}
\PrintBackRefs{\CurrentBib}

\bibitem [\protect \citeauthoryear {%
{Natural Language Toolkit (NLTK) Project}%
}{%
{Natural Language Toolkit (NLTK) Project}%
}{%
{\protect \APACyear {2022}}%
}]{%
nltk}
\APACinsertmetastar {%
nltk}%
\begin{APACrefauthors}%
{Natural Language Toolkit (NLTK) Project}%
\end{APACrefauthors}%
\unskip\
\newblock
\APACrefYearMonthDay{2022}{}{}.
\newblock
\APACrefbtitle {{Source code for \texttt{nltk.metrics.agreement}}.} {{Source
  code for \texttt{nltk.metrics.agreement}}.}
\newblock
\APAChowpublished {Retrieved April 6, 2022 from
  \url{http://www.nltk.org/\_modules/nltk/metrics/agreement.html}}.
\PrintBackRefs{\CurrentBib}

\bibitem [\protect \citeauthoryear {%
Nunn%
, Avella%
, Kanai%
\BCBL {}\ \BBA {} Kebritchi%
}{%
Nunn%
\ \protect \BOthers {.}}{%
{\protect \APACyear {2016}}%
}]{%
avella2016}
\APACinsertmetastar {%
avella2016}%
\begin{APACrefauthors}%
Nunn, S.G.%
, Avella, J.T.%
, Kanai, T.%
\BCBL {} Kebritchi, M.%
\end{APACrefauthors}%
\unskip\
\newblock
\APACrefYearMonthDay{2016}{}{}.
\newblock
{\BBOQ}\APACrefatitle {{Learning Analytics Methods, Benefits, and Challenges in
  Higher Education: A Systematic Literature Review}} {{Learning Analytics
  Methods, Benefits, and Challenges in Higher Education: A Systematic
  Literature Review}}.{\BBCQ}
\newblock
\APACjournalVolNumPages{Online Learning}{20}{2}{13--29}.
\newblock
\begin{APACrefURL} {\url{https://doi.org/10.24059/olj.v20i2.790}}
  \end{APACrefURL}
\newblock

\newblock

\PrintBackRefs{\CurrentBib}

\bibitem [\protect \citeauthoryear {%
{Palmer}%
}{%
{Palmer}%
}{%
{\protect \APACyear {2019}}%
}]{%
Palmer2019}
\APACinsertmetastar {%
Palmer2019}%
\begin{APACrefauthors}%
{Palmer}, N.%
\end{APACrefauthors}%
\unskip\
\newblock
\APACrefYearMonthDay{2019}{}{}.
\newblock
{\BBOQ}\APACrefatitle {{Automating the Assessment of Network Security in Higher
  Education}} {{Automating the Assessment of Network Security in Higher
  Education}}.{\BBCQ}
\newblock
 \APACrefbtitle {{2019 International Conference on Computing, Electronics
  Communications Engineering (iCCECE)}} {{2019 International Conference on
  Computing, Electronics Communications Engineering (iCCECE)}}\ (\BPGS\
  141--146).
\newblock
\begin{APACrefURL} {\url{https://doi.org/10.1109/iCCECE46942.2019.8941804}}
  \end{APACrefURL}
\PrintBackRefs{\CurrentBib}

\bibitem [\protect \citeauthoryear {%
Papamitsiou%
, Giannakos%
, Simon%
\BCBL {}\ \BBA {} Luxton-Reilly%
}{%
Papamitsiou%
\ \protect \BOthers {.}}{%
{\protect \APACyear {2020}}%
}]{%
Papamitsiou2020}
\APACinsertmetastar {%
Papamitsiou2020}%
\begin{APACrefauthors}%
Papamitsiou, Z.%
, Giannakos, M.%
, Simon%
\BCBL {} Luxton-Reilly, A.%
\end{APACrefauthors}%
\unskip\
\newblock
\APACrefYearMonthDay{2020}{}{}.
\newblock
{\BBOQ}\APACrefatitle {{Computing Education Research Landscape through an
  Analysis of Keywords}} {{Computing Education Research Landscape through an
  Analysis of Keywords}}.{\BBCQ}
\newblock
 \APACrefbtitle {{Proceedings of the 2020 ACM Conference on International
  Computing Education Research}} {{Proceedings of the 2020 ACM Conference on
  International Computing Education Research}}\ (\BPGS\ 102--112).
\newblock
\APACaddressPublisher{New York, NY, USA}{Association for Computing Machinery}.
\newblock
\begin{APACrefURL} {\url{https://doi.org/10.1145/3372782.3406276}}
  \end{APACrefURL}
\PrintBackRefs{\CurrentBib}

\bibitem [\protect \citeauthoryear {%
Petersen%
, Feldt%
, Mujtaba%
\BCBL {}\ \BBA {} Mattsson%
}{%
Petersen%
\ \protect \BOthers {.}}{%
{\protect \APACyear {2008}}%
}]{%
Petersen:2008}
\APACinsertmetastar {%
Petersen:2008}%
\begin{APACrefauthors}%
Petersen, K.%
, Feldt, R.%
, Mujtaba, S.%
\BCBL {} Mattsson, M.%
\end{APACrefauthors}%
\unskip\
\newblock
\APACrefYearMonthDay{2008}{}{}.
\newblock
{\BBOQ}\APACrefatitle {{Systematic Mapping Studies in Software Engineering}}
  {{Systematic Mapping Studies in Software Engineering}}.{\BBCQ}
\newblock
 \APACrefbtitle {{Proceedings of the 12th International Conference on
  Evaluation and Assessment in Software Engineering}} {{Proceedings of the 12th
  International Conference on Evaluation and Assessment in Software
  Engineering}}\ (\BPGS\ 68--77).
\newblock
\APACaddressPublisher{Swindon, UK}{BCS Learning \& Development Ltd.}
\newblock
\begin{APACrefURL} {\url{https://dl.acm.org/doi/10.5555/2227115.2227123}}
  \end{APACrefURL}
\PrintBackRefs{\CurrentBib}

\bibitem [\protect \citeauthoryear {%
Petersen%
, Vakkalanka%
\BCBL {}\ \BBA {} Kuzniarz%
}{%
Petersen%
\ \protect \BOthers {.}}{%
{\protect \APACyear {2015}}%
}]{%
petersen2015}
\APACinsertmetastar {%
petersen2015}%
\begin{APACrefauthors}%
Petersen, K.%
, Vakkalanka, S.%
\BCBL {} Kuzniarz, L.%
\end{APACrefauthors}%
\unskip\
\newblock
\APACrefYearMonthDay{2015}{}{}.
\newblock
{\BBOQ}\APACrefatitle {Guidelines for conducting systematic mapping studies in
  software engineering: An update} {Guidelines for conducting systematic
  mapping studies in software engineering: An update}.{\BBCQ}
\newblock
\APACjournalVolNumPages{Information and Software Technology}{64}{}{1--18}.
\newblock
\begin{APACrefURL} {\url{https://doi.org/10.1016/j.infsof.2015.03.007}}
  \end{APACrefURL}
\newblock

\newblock

\PrintBackRefs{\CurrentBib}

\bibitem [\protect \citeauthoryear {%
Peña-Ayala%
}{%
Peña-Ayala%
}{%
{\protect \APACyear {2014}}%
}]{%
penaayala2014}
\APACinsertmetastar {%
penaayala2014}%
\begin{APACrefauthors}%
Peña-Ayala, A.%
\end{APACrefauthors}%
\unskip\
\newblock
\APACrefYearMonthDay{2014}{}{}.
\newblock
{\BBOQ}\APACrefatitle {Educational data mining: A survey and a data
  mining-based analysis of recent works} {Educational data mining: A survey and
  a data mining-based analysis of recent works}.{\BBCQ}
\newblock
\APACjournalVolNumPages{Expert Systems with Applications}{41}{4, Part
  1}{1432--1462}.
\newblock
\begin{APACrefURL} {\url{https://doi.org/10.1016/j.eswa.2013.08.042}}
  \end{APACrefURL}
\newblock

\newblock

\PrintBackRefs{\CurrentBib}

\bibitem [\protect \citeauthoryear {%
Randolph%
}{%
Randolph%
}{%
{\protect \APACyear {2009}}%
}]{%
randolph2009guide}
\APACinsertmetastar {%
randolph2009guide}%
\begin{APACrefauthors}%
Randolph, J.%
\end{APACrefauthors}%
\unskip\
\newblock
\APACrefYearMonthDay{2009}{}{}.
\newblock
{\BBOQ}\APACrefatitle {A guide to writing the dissertation literature review}
  {A guide to writing the dissertation literature review}.{\BBCQ}
\newblock
\APACjournalVolNumPages{Practical Assessment, Research, and
  Evaluation}{14}{1}{13}.
\newblock
\begin{APACrefURL} {\url{https://doi.org/10.7275/b0az-8t74}} \end{APACrefURL}
\newblock

\newblock

\PrintBackRefs{\CurrentBib}

\bibitem [\protect \citeauthoryear {%
Reed%
, Nauer%
\BCBL {}\ \BBA {} Silva%
}{%
Reed%
\ \protect \BOthers {.}}{%
{\protect \APACyear {2013}}%
}]{%
reed2013instrumenting}
\APACinsertmetastar {%
reed2013instrumenting}%
\begin{APACrefauthors}%
Reed, T.%
, Nauer, K.%
\BCBL {} Silva, A.%
\end{APACrefauthors}%
\unskip\
\newblock
\APACrefYearMonthDay{2013}{}{}.
\newblock
{\BBOQ}\APACrefatitle {Instrumenting competition-based exercises to evaluate
  cyber defender situation awareness} {Instrumenting competition-based
  exercises to evaluate cyber defender situation awareness}.{\BBCQ}
\newblock
 \APACrefbtitle {{International Conference on Augmented Cognition}}
  {{International Conference on Augmented Cognition}}\ (\BPGS\ 80--89).
\newblock
\APACaddressPublisher{Vienna, Austria}{Springer}.
\newblock
\begin{APACrefURL} {\url{https://doi.org/10.1007/978-3-642-39454-6_9}}
  \end{APACrefURL}
\PrintBackRefs{\CurrentBib}

\bibitem [\protect \citeauthoryear {%
Rege%
\ \protect \BOthers {.}}{%
Rege%
\ \protect \BOthers {.}}{%
{\protect \APACyear {2017}}%
}]{%
Rege2017}
\APACinsertmetastar {%
Rege2017}%
\begin{APACrefauthors}%
Rege, A.%
, Obradovic, Z.%
, Asadi, N.%
, Parker, E.%
, Masceri, N.%
, Singer, B.%
\BCBL {} Pandit, R.%
\end{APACrefauthors}%
\unskip\
\newblock
\APACrefYearMonthDay{2017}{}{}.
\newblock
{\BBOQ}\APACrefatitle {{Using a Real-Time Cybersecurity Exercise Case Study to
  Understand Temporal Characteristics of Cyberattacks}} {{Using a Real-Time
  Cybersecurity Exercise Case Study to Understand Temporal Characteristics of
  Cyberattacks}}.{\BBCQ}
\newblock
 \APACrefbtitle {Social, Cultural, and Behavioral Modeling} {Social, cultural,
  and behavioral modeling}\ (\BPGS\ 127--132).
\newblock
\APACaddressPublisher{Cham, Switzerland}{Springer International Publishing}.
\newblock
\begin{APACrefURL} {\url{https://doi.org/10.1007/978-3-319-60240-0_16}}
  \end{APACrefURL}
\PrintBackRefs{\CurrentBib}

\bibitem [\protect \citeauthoryear {%
Romero%
, Ventura%
, Pechenizkiy%
\BCBL {}\ \BBA {} Baker%
}{%
Romero%
\ \protect \BOthers {.}}{%
{\protect \APACyear {2010}}%
}]{%
handbook-edm2010}
\APACinsertmetastar {%
handbook-edm2010}%
\begin{APACrefauthors}%
Romero, C.%
, Ventura, S.%
, Pechenizkiy, M.%
\BCBL {}\ \BBA {} Baker, R.S.%
\end{APACrefauthors}%
\ (\BEDS).
\unskip\
\newblock
\APACrefYear{2010}.
\newblock
\APACrefbtitle {Handbook of educational data mining} {Handbook of educational
  data mining}.
\newblock
\APACaddressPublisher{Boca Raton, FL, USA}{CRC Press}.
\newblock
\begin{APACrefURL} {\url{https://doi.org/10.1201/b10274}} \end{APACrefURL}
\PrintBackRefs{\CurrentBib}

\bibitem [\protect \citeauthoryear {%
Rupp%
\ \protect \BOthers {.}}{%
Rupp%
\ \protect \BOthers {.}}{%
{\protect \APACyear {2012}}%
}]{%
rupp2012putting}
\APACinsertmetastar {%
rupp2012putting}%
\begin{APACrefauthors}%
Rupp, A.A.%
, Levy, R.%
, Dicerbo, K.E.%
, Sweet, S.J.%
, Crawford, A.V.%
, Calico, T.%
\BDBL {}Behrens, J.T.%
\end{APACrefauthors}%
\unskip\
\newblock
\APACrefYearMonthDay{2012}{}{}.
\newblock
{\BBOQ}\APACrefatitle {{Putting ECD into practice: The interplay of theory and
  data in evidence models within a digital learning environment}} {{Putting ECD
  into practice: The interplay of theory and data in evidence models within a
  digital learning environment}}.{\BBCQ}
\newblock
\APACjournalVolNumPages{Journal of Educational Data Mining
  (JEDM)}{4}{1}{49--110}.
\newblock
\begin{APACrefURL} {\url{https://doi.org/10.5281/zenodo.3554643}}
  \end{APACrefURL}
\newblock

\newblock

\PrintBackRefs{\CurrentBib}

\bibitem [\protect \citeauthoryear {%
Sanders%
\ \protect \BOthers {.}}{%
Sanders%
\ \protect \BOthers {.}}{%
{\protect \APACyear {2019}}%
}]{%
Sanders2019}
\APACinsertmetastar {%
Sanders2019}%
\begin{APACrefauthors}%
Sanders, K.%
, Sheard, J.%
, Becker, B.A.%
, Eckerdal, A.%
, Hamouda, S.%
\BCBL {} Simon.%
\end{APACrefauthors}%
\unskip\
\newblock
\APACrefYearMonthDay{2019}{}{}.
\newblock
{\BBOQ}\APACrefatitle {{Inferential Statistics in Computing Education Research:
  A Methodological Review}} {{Inferential Statistics in Computing Education
  Research: A Methodological Review}}.{\BBCQ}
\newblock
 \APACrefbtitle {{Proceedings of the 2019 ACM Conference on International
  Computing Education Research}} {{Proceedings of the 2019 ACM Conference on
  International Computing Education Research}}\ (\BPG~177–185).
\newblock
\APACaddressPublisher{New York, NY, USA}{Association for Computing Machinery}.
\newblock
\begin{APACrefURL} {\url{https://doi.org/10.1145/3291279.3339408}}
  \end{APACrefURL}
\PrintBackRefs{\CurrentBib}

\bibitem [\protect \citeauthoryear {%
Sheng%
}{%
Sheng%
}{%
{\protect \APACyear {2020}}%
}]{%
Sheng2020}
\APACinsertmetastar {%
Sheng2020}%
\begin{APACrefauthors}%
Sheng, Q\BHBI w.%
\end{APACrefauthors}%
\unskip\
\newblock
\APACrefYearMonthDay{2020}{}{}.
\newblock
{\BBOQ}\APACrefatitle {{Effectiveness Evaluation of Network Security Knowledge
  Training Based on Machine Learning}} {{Effectiveness Evaluation of Network
  Security Knowledge Training Based on Machine Learning}}.{\BBCQ}
\newblock
 S.~Liu, G.~Sun\BCBL {}\ \BBA {} W.~Fu\ (\BEDS), \APACrefbtitle {{e-Learning,
  e-Education, and Online Training}} {{e-Learning, e-Education, and Online
  Training}}\ (\BPGS\ 25--37).
\newblock
\APACaddressPublisher{Cham}{Springer International Publishing}.
\newblock
\begin{APACrefURL} {\url{https://doi.org/10.1007/978-3-030-63955-6_3}}
  \end{APACrefURL}
\PrintBackRefs{\CurrentBib}

\bibitem [\protect \citeauthoryear {%
Simon%
\ \protect \BOthers {.}}{%
Simon%
\ \protect \BOthers {.}}{%
{\protect \APACyear {2019}}%
}]{%
Simon2019}
\APACinsertmetastar {%
Simon2019}%
\begin{APACrefauthors}%
Simon, S.%
, Becker, B.A.%
, Hamouda, S.%
, McCartney, R.%
, Sanders, K.%
\BCBL {} Sheard, J.%
\end{APACrefauthors}%
\unskip\
\newblock
\APACrefYearMonthDay{2019}{}{}.
\newblock
{\BBOQ}\APACrefatitle {{Visual Portrayals of Data and Results at ITiCSE}}
  {{Visual Portrayals of Data and Results at ITiCSE}}.{\BBCQ}
\newblock
 \APACrefbtitle {{Proceedings of the 2019 ACM Conference on Innovation and
  Technology in Computer Science Education}} {{Proceedings of the 2019 ACM
  Conference on Innovation and Technology in Computer Science Education}}\
  (\BPG~51–57).
\newblock
\APACaddressPublisher{New York, NY, USA}{Association for Computing Machinery}.
\newblock
\begin{APACrefURL} {\url{https://doi.org/10.1145/3304221.3319742}}
  \end{APACrefURL}
\PrintBackRefs{\CurrentBib}

\bibitem [\protect \citeauthoryear {%
Taylor%
, Arias%
, Klopchic%
, Matarazzo%
\BCBL {}\ \BBA {} Dube%
}{%
Taylor%
\ \protect \BOthers {.}}{%
{\protect \APACyear {2017}}%
}]{%
taylor2017}
\APACinsertmetastar {%
taylor2017}%
\begin{APACrefauthors}%
Taylor, C.%
, Arias, P.%
, Klopchic, J.%
, Matarazzo, C.%
\BCBL {} Dube, E.%
\end{APACrefauthors}%
\unskip\
\newblock
\APACrefYearMonthDay{2017}{}{}.
\newblock
{\BBOQ}\APACrefatitle {{CTF}: State-of-the-Art and Building the Next
  Generation} {{CTF}: State-of-the-art and building the next
  generation}.{\BBCQ}
\newblock
 \APACrefbtitle {{2017 {USENIX} Workshop on Advances in Security Education
  ({ASE} 17)}.} {{2017 {USENIX} Workshop on Advances in Security Education
  ({ASE} 17)}.}
\newblock
\APACaddressPublisher{}{{USENIX} Association}.
\newblock
\begin{APACrefURL}
  {\url{https://www.usenix.org/conference/ase17/workshop-program/presentation/taylor}}
  \end{APACrefURL}
\PrintBackRefs{\CurrentBib}

\bibitem [\protect \citeauthoryear {%
{The Graphics Replicability Stamp Initiative}%
}{%
{The Graphics Replicability Stamp Initiative}%
}{%
{\protect \APACyear {2017}}%
}]{%
grsi}
\APACinsertmetastar {%
grsi}%
\begin{APACrefauthors}%
{The Graphics Replicability Stamp Initiative}%
\end{APACrefauthors}%
\unskip\
\newblock
\APACrefYearMonthDay{2017}{}{}.
\newblock
\APACrefbtitle {{GRSI}.} {{GRSI}.}
\newblock
\APAChowpublished {Retrieved April 6, 2022 from
  \url{http://www.replicabilitystamp.org}}.
\PrintBackRefs{\CurrentBib}

\bibitem [\protect \citeauthoryear {%
{The NATO Cooperative Cyber Defence Centre of Excellence}%
}{%
{The NATO Cooperative Cyber Defence Centre of Excellence}%
}{%
{\protect \APACyear {2021}}%
{\protect \APACexlab {{\protect \BCnt {1}}}}}]{%
crossed-swords}
\APACinsertmetastar {%
crossed-swords}%
\begin{APACrefauthors}%
{The NATO Cooperative Cyber Defence Centre of Excellence}%
\end{APACrefauthors}%
\unskip\
\newblock
\APACrefYearMonthDay{2021{\protect \BCnt {1}}}{}{}.
\newblock
\APACrefbtitle {{Crossed Swords}.} {{Crossed Swords}.}
\newblock
\APAChowpublished {Retrieved April 6, 2022 from
  \url{https://ccdcoe.org/exercises/crossed-swords}}.
\PrintBackRefs{\CurrentBib}

\bibitem [\protect \citeauthoryear {%
{The NATO Cooperative Cyber Defence Centre of Excellence}%
}{%
{The NATO Cooperative Cyber Defence Centre of Excellence}%
}{%
{\protect \APACyear {2021}}%
{\protect \APACexlab {{\protect \BCnt {2}}}}}]{%
locked-shields}
\APACinsertmetastar {%
locked-shields}%
\begin{APACrefauthors}%
{The NATO Cooperative Cyber Defence Centre of Excellence}%
\end{APACrefauthors}%
\unskip\
\newblock
\APACrefYearMonthDay{2021{\protect \BCnt {2}}}{}{}.
\newblock
\APACrefbtitle {{Locked Shields}.} {{Locked Shields}.}
\newblock
\APAChowpublished {Retrieved April 6, 2022 from
  \url{https://ccdcoe.org/exercises/locked-shields}}.
\PrintBackRefs{\CurrentBib}

\bibitem [\protect \citeauthoryear {%
Tian%
\ \protect \BOthers {.}}{%
Tian%
\ \protect \BOthers {.}}{%
{\protect \APACyear {2018}}%
}]{%
Tian2018}
\APACinsertmetastar {%
Tian2018}%
\begin{APACrefauthors}%
Tian, Z.%
, Cui, Y.%
, An, L.%
, Su, S.%
, Yin, X.%
, Yin, L.%
\BCBL {} Cui, X.%
\end{APACrefauthors}%
\unskip\
\newblock
\APACrefYearMonthDay{2018}{}{}.
\newblock
{\BBOQ}\APACrefatitle {{A real-time correlation of host-level events in cyber
  range service for smart campus}} {{A real-time correlation of host-level
  events in cyber range service for smart campus}}.{\BBCQ}
\newblock
\APACjournalVolNumPages{IEEE Access}{6}{}{35355--35364}.
\newblock
\begin{APACrefURL} {\url{https://doi.org/10.1109/ACCESS.2018.2846590}}
  \end{APACrefURL}
\newblock

\newblock

\PrintBackRefs{\CurrentBib}

\bibitem [\protect \citeauthoryear {%
Tobarra%
\ \protect \BOthers {.}}{%
Tobarra%
\ \protect \BOthers {.}}{%
{\protect \APACyear {2020}}%
}]{%
Tobarra2020-acceptance}
\APACinsertmetastar {%
Tobarra2020-acceptance}%
\begin{APACrefauthors}%
Tobarra, L.%
, Robles-Gómez, A.%
, Pastor, R.%
, Hernández, R.%
, Duque, A.%
\BCBL {} Cano, J.%
\end{APACrefauthors}%
\unskip\
\newblock
\APACrefYearMonthDay{2020}{02}{}.
\newblock
{\BBOQ}\APACrefatitle {Students’ Acceptance and Tracking of a New
  Container-Based Virtual Laboratory} {Students’ acceptance and tracking of a
  new container-based virtual laboratory}.{\BBCQ}
\newblock
\APACjournalVolNumPages{Applied Sciences}{10}{3}{}.
\newblock
\begin{APACrefURL} {\url{https://doi.org/10.3390/app10031091}} \end{APACrefURL}
\newblock

\newblock

\PrintBackRefs{\CurrentBib}

\bibitem [\protect \citeauthoryear {%
{Tobarra}%
\ \protect \BOthers {.}}{%
{Tobarra}%
\ \protect \BOthers {.}}{%
{\protect \APACyear {2020}}%
}]{%
Tobarra2020-game}
\APACinsertmetastar {%
Tobarra2020-game}%
\begin{APACrefauthors}%
{Tobarra}, L.%
, {Trapero}, A.P.%
, {Pastor}, R.%
, {Robles-Gómez}, A.%
, {Hernández}, R.%
, {Duque}, A.%
\BCBL {} {Cano}, J.%
\end{APACrefauthors}%
\unskip\
\newblock
\APACrefYearMonthDay{2020}{}{}.
\newblock
{\BBOQ}\APACrefatitle {Game-based Learning Approach to Cybersecurity}
  {Game-based learning approach to cybersecurity}.{\BBCQ}
\newblock
 \APACrefbtitle {{2020 IEEE Global Engineering Education Conference (EDUCON)}}
  {{2020 IEEE Global Engineering Education Conference (EDUCON)}}\
  (\BPG~1125-1132).
\newblock
\begin{APACrefURL} {\url{https://doi.org/10.1109/EDUCON45650.2020.9125202}}
  \end{APACrefURL}
\PrintBackRefs{\CurrentBib}

\bibitem [\protect \citeauthoryear {%
{Tseng}%
\ \protect \BOthers {.}}{%
{Tseng}%
\ \protect \BOthers {.}}{%
{\protect \APACyear {2017}}%
}]{%
Tseng2017ontology}
\APACinsertmetastar {%
Tseng2017ontology}%
\begin{APACrefauthors}%
{Tseng}, S\BHBI S.%
, {Lin}, S\BHBI C.%
, {Mao}, C\BHBI H.%
, {Lee}, T\BHBI J.%
, {Qiu}, G\BHBI W.%
\BCBL {} {Lin}, M\BHBI H.%
\end{APACrefauthors}%
\unskip\
\newblock
\APACrefYearMonthDay{2017}{08}{}.
\newblock
{\BBOQ}\APACrefatitle {An ontology guiding assessment framework for hacking
  competition} {An ontology guiding assessment framework for hacking
  competition}.{\BBCQ}
\newblock
 \APACrefbtitle {{2017 10th International Conference on Ubi-media Computing and
  Workshops (Ubi-Media)}} {{2017 10th International Conference on Ubi-media
  Computing and Workshops (Ubi-Media)}}\ (\BPGS\ 1--4).
\newblock
\APACaddressPublisher{New York, NY, USA}{IEEE}.
\newblock
\begin{APACrefURL} {\url{https://doi.org/10.1109/UMEDIA.2017.8074131}}
  \end{APACrefURL}
\PrintBackRefs{\CurrentBib}

\bibitem [\protect \citeauthoryear {%
Vigna%
\ \protect \BOthers {.}}{%
Vigna%
\ \protect \BOthers {.}}{%
{\protect \APACyear {2014}}%
}]{%
vigna2014}
\APACinsertmetastar {%
vigna2014}%
\begin{APACrefauthors}%
Vigna, G.%
, Borgolte, K.%
, Corbetta, J.%
, Doup{\'e}, A.%
, Fratantonio, Y.%
, Invernizzi, L.%
\BDBL {}Shoshitaishvili, Y.%
\end{APACrefauthors}%
\unskip\
\newblock
\APACrefYearMonthDay{2014}{}{}.
\newblock
{\BBOQ}\APACrefatitle {{Ten Years of iCTF: The Good, The Bad, and The Ugly}}
  {{Ten Years of iCTF: The Good, The Bad, and The Ugly}}.{\BBCQ}
\newblock
 \APACrefbtitle {{2014 {USENIX} Summit on Gaming, Games, and Gamification in
  Security Education (3GSE 14)}} {{2014 {USENIX} Summit on Gaming, Games, and
  Gamification in Security Education (3GSE 14)}}\ (\BPGS\ 1--7).
\newblock
\APACaddressPublisher{San Diego, CA}{{USENIX} Association}.
\newblock
\begin{APACrefURL}
  {\url{https://www.usenix.org/conference/3gse14/summit-program/presentation/vigna}}
  \end{APACrefURL}
\PrintBackRefs{\CurrentBib}

\bibitem [\protect \citeauthoryear {%
\v{S}v\'{a}bensk\'{y}%
\ \BBA {} Vykopal%
}{%
\v{S}v\'{a}bensk\'{y}%
\ \BBA {} Vykopal%
}{%
{\protect \APACyear {2018}}%
{\protect \APACexlab {{\protect \BCnt {1}}}}}]{%
Svabensky2018challenges}
\APACinsertmetastar {%
Svabensky2018challenges}%
\begin{APACrefauthors}%
\v{S}v\'{a}bensk\'{y}, V.%
\BCBT {}\ \BBA {} Vykopal, J.%
\end{APACrefauthors}%
\unskip\
\newblock
\APACrefYearMonthDay{2018{\protect \BCnt {1}}}{02}{}.
\newblock
{\BBOQ}\APACrefatitle {{Challenges Arising from Prerequisite Testing in
  Cybersecurity Games}} {{Challenges Arising from Prerequisite Testing in
  Cybersecurity Games}}.{\BBCQ}
\newblock
 \APACrefbtitle {{Proceedings of the 49th ACM Technical Symposium on Computer
  Science Education}} {{Proceedings of the 49th ACM Technical Symposium on
  Computer Science Education}}\ (\BPGS\ 56--61).
\newblock
\APACaddressPublisher{New York, NY, USA}{Association for Computing Machinery}.
\newblock
\begin{APACrefURL} {\url{https://doi.org/10.1145/3159450.3159454}}
  \end{APACrefURL}
\PrintBackRefs{\CurrentBib}

\bibitem [\protect \citeauthoryear {%
\v{S}v\'{a}bensk\'{y}%
\ \BBA {} Vykopal%
}{%
\v{S}v\'{a}bensk\'{y}%
\ \BBA {} Vykopal%
}{%
{\protect \APACyear {2018}}%
{\protect \APACexlab {{\protect \BCnt {2}}}}}]{%
Svabensky2018gathering}
\APACinsertmetastar {%
Svabensky2018gathering}%
\begin{APACrefauthors}%
\v{S}v\'{a}bensk\'{y}, V.%
\BCBT {}\ \BBA {} Vykopal, J.%
\end{APACrefauthors}%
\unskip\
\newblock
\APACrefYearMonthDay{2018{\protect \BCnt {2}}}{10}{}.
\newblock
{\BBOQ}\APACrefatitle {{Gathering Insights from Teenagers' Hacking Experience
  with Authentic Cybersecurity Tools}} {{Gathering Insights from Teenagers'
  Hacking Experience with Authentic Cybersecurity Tools}}.{\BBCQ}
\newblock
 \APACrefbtitle {{Proceedings of the 48th IEEE Frontiers in Education
  Conference}} {{Proceedings of the 48th IEEE Frontiers in Education
  Conference}}\ (\BPGS\ 1--4).
\newblock
\APACaddressPublisher{New York, NY, USA}{IEEE}.
\newblock
\begin{APACrefURL} {\url{https://doi.org/10.1109/FIE.2018.8658840}}
  \end{APACrefURL}
\PrintBackRefs{\CurrentBib}

\bibitem [\protect \citeauthoryear {%
\v{S}v\'{a}bensk\'{y}%
, Vykopal%
\BCBL {}\ \BBA {} \v{C}eleda%
}{%
\v{S}v\'{a}bensk\'{y}%
, Vykopal%
\BCBL {}\ \BBA {} \v{C}eleda%
}{%
{\protect \APACyear {2020}}%
}]{%
Svabensky2020}
\APACinsertmetastar {%
Svabensky2020}%
\begin{APACrefauthors}%
\v{S}v\'{a}bensk\'{y}, V.%
, Vykopal, J.%
\BCBL {} \v{C}eleda, P.%
\end{APACrefauthors}%
\unskip\
\newblock
\APACrefYearMonthDay{2020}{03}{}.
\newblock
{\BBOQ}\APACrefatitle {{What Are Cybersecurity Education Papers About? A
  Systematic Literature Review of SIGCSE and ITiCSE Conferences}} {{What Are
  Cybersecurity Education Papers About? A Systematic Literature Review of
  SIGCSE and ITiCSE Conferences}}.{\BBCQ}
\newblock
 \APACrefbtitle {{Proceedings of the 51st ACM Technical Symposium on Computer
  Science Education}} {{Proceedings of the 51st ACM Technical Symposium on
  Computer Science Education}}\ (\BPGS\ 2--8).
\newblock
\APACaddressPublisher{New York, NY, USA}{Association for Computing Machinery}.
\newblock
\begin{APACrefURL} {\url{https://doi.org/10.1145/3328778.3366816}}
  \end{APACrefURL}
\PrintBackRefs{\CurrentBib}

\bibitem [\protect \citeauthoryear {%
\v{S}v\'{a}bensk\'{y}%
, Vykopal%
, \v{C}eleda%
\BCBL {}\ \BBA {} Kraus%
}{%
\v{S}v\'{a}bensk\'{y}%
\ \protect \BOthers {.}}{%
{\protect \APACyear {2022}}%
}]{%
supplementary-materials}
\APACinsertmetastar {%
supplementary-materials}%
\begin{APACrefauthors}%
\v{S}v\'{a}bensk\'{y}, V.%
, Vykopal, J.%
, \v{C}eleda, P.%
\BCBL {} Kraus, L.%
\end{APACrefauthors}%
\unskip\
\newblock
\APACrefYearMonthDay{2022}{}{}.
\newblock
\APACrefbtitle {{Dataset: Applications of Educational Data Mining and Learning
  Analytics on Data From Cybersecurity Training}.} {{Dataset: Applications of
  Educational Data Mining and Learning Analytics on Data From Cybersecurity
  Training}.}
\newblock
\APACaddressPublisher{}{Zenodo}.
\newblock
\begin{APACrefURL} {\url{https://doi.org/10.5281/zenodo.6573117}}
  \end{APACrefURL}
\PrintBackRefs{\CurrentBib}

\bibitem [\protect \citeauthoryear {%
\v{S}v\'{a}bensk\'{y}%
, Čeleda%
, Vykopal%
\BCBL {}\ \BBA {} Brišáková%
}{%
\v{S}v\'{a}bensk\'{y}%
, Čeleda%
\BCBL {}\ \protect \BOthers {.}}{%
{\protect \APACyear {2020}}%
}]{%
Svabensky2021}
\APACinsertmetastar {%
Svabensky2021}%
\begin{APACrefauthors}%
\v{S}v\'{a}bensk\'{y}, V.%
, Čeleda, P.%
, Vykopal, J.%
\BCBL {} Brišáková, S.%
\end{APACrefauthors}%
\unskip\
\newblock
\APACrefYearMonthDay{2020}{12}{}.
\newblock
{\BBOQ}\APACrefatitle {{Cybersecurity Knowledge and Skills Taught in Capture
  the Flag Challenges}} {{Cybersecurity Knowledge and Skills Taught in Capture
  the Flag Challenges}}.{\BBCQ}
\newblock
\APACjournalVolNumPages{Elsevier Computers~\& Security}{102}{102154}{}.
\newblock
\begin{APACrefURL} {\url{https://doi.org/10.1016/j.cose.2020.1021548}}
  \end{APACrefURL}
\newblock

\newblock

\PrintBackRefs{\CurrentBib}

\bibitem [\protect \citeauthoryear {%
Vykopal%
\ \BBA {} Bart{\'a}k%
}{%
Vykopal%
\ \BBA {} Bart{\'a}k%
}{%
{\protect \APACyear {2016}}%
}]{%
Vykopal2016}
\APACinsertmetastar {%
Vykopal2016}%
\begin{APACrefauthors}%
Vykopal, J.%
\BCBT {}\ \BBA {} Bart{\'a}k, M.%
\end{APACrefauthors}%
\unskip\
\newblock
\APACrefYearMonthDay{2016}{}{}.
\newblock
{\BBOQ}\APACrefatitle {{On the Design of Security Games: From Frustrating to
  Engaging Learning}} {{On the Design of Security Games: From Frustrating to
  Engaging Learning}}.{\BBCQ}
\newblock
 \APACrefbtitle {{2016 {USENIX} Workshop on Advances in Security Education
  ({ASE} 16)}} {{2016 {USENIX} Workshop on Advances in Security Education
  ({ASE} 16)}}\ (\BPGS\ 1--8).
\newblock
\APACaddressPublisher{Berkeley, CA, USA}{{USENIX} Association}.
\newblock
\begin{APACrefURL}
  {\url{https://www.usenix.org/conference/ase16/workshop-program/presentation/vykopal}}
  \end{APACrefURL}
\PrintBackRefs{\CurrentBib}

\bibitem [\protect \citeauthoryear {%
{Vykopal}%
, {Vizvary}%
, {Oslejsek}%
, {Celeda}%
\BCBL {}\ \BBA {} {Tovarnak}%
}{%
{Vykopal}%
\ \protect \BOthers {.}}{%
{\protect \APACyear {2017}}%
}]{%
Vykopal2017}
\APACinsertmetastar {%
Vykopal2017}%
\begin{APACrefauthors}%
{Vykopal}, J.%
, {Vizvary}, M.%
, {Oslejsek}, R.%
, {Celeda}, P.%
\BCBL {} {Tovarnak}, D.%
\end{APACrefauthors}%
\unskip\
\newblock
\APACrefYearMonthDay{2017}{}{}.
\newblock
{\BBOQ}\APACrefatitle {Lessons learned from complex hands-on defence exercises
  in a cyber range} {Lessons learned from complex hands-on defence exercises in
  a cyber range}.{\BBCQ}
\newblock
 \APACrefbtitle {{2017 IEEE Frontiers in Education Conference (FIE)}} {{2017
  IEEE Frontiers in Education Conference (FIE)}}\ (\BPGS\ 1--8).
\newblock
\APACaddressPublisher{New York, NY, USA}{IEEE}.
\newblock
\begin{APACrefURL} {\url{https://doi.org/10.1109/FIE.2017.8190713}}
  \end{APACrefURL}
\PrintBackRefs{\CurrentBib}

\bibitem [\protect \citeauthoryear {%
Weiss%
, Locasto%
\BCBL {}\ \BBA {} Mache%
}{%
Weiss%
\ \protect \BOthers {.}}{%
{\protect \APACyear {2016}}%
}]{%
Weiss2016reflective}
\APACinsertmetastar {%
Weiss2016reflective}%
\begin{APACrefauthors}%
Weiss, R.%
, Locasto, M.E.%
\BCBL {} Mache, J.%
\end{APACrefauthors}%
\unskip\
\newblock
\APACrefYearMonthDay{2016}{}{}.
\newblock
{\BBOQ}\APACrefatitle {{A Reflective Approach to Assessing Student Performance
  in Cybersecurity Exercises}} {{A Reflective Approach to Assessing Student
  Performance in Cybersecurity Exercises}}.{\BBCQ}
\newblock
 \APACrefbtitle {{Proceedings of the 47th ACM Technical Symposium on Computing
  Science Education}} {{Proceedings of the 47th ACM Technical Symposium on
  Computing Science Education}}\ (\BPGS\ 597--602).
\newblock
\APACaddressPublisher{New York, NY, USA}{ACM}.
\newblock
\begin{APACrefURL} {\url{https://doi.org/10.1145/2839509.2844646}}
  \end{APACrefURL}
\PrintBackRefs{\CurrentBib}

\bibitem [\protect \citeauthoryear {%
Weiss%
, Turbak%
, Mache%
\BCBL {}\ \BBA {} Locasto%
}{%
Weiss%
\ \protect \BOthers {.}}{%
{\protect \APACyear {2017}}%
}]{%
Weiss2017}
\APACinsertmetastar {%
Weiss2017}%
\begin{APACrefauthors}%
Weiss, R.%
, Turbak, F.%
, Mache, J.%
\BCBL {} Locasto, M.E.%
\end{APACrefauthors}%
\unskip\
\newblock
\APACrefYearMonthDay{2017}{}{}.
\newblock
{\BBOQ}\APACrefatitle {{Cybersecurity Education and Assessment in EDURange}}
  {{Cybersecurity Education and Assessment in EDURange}}.{\BBCQ}
\newblock
\APACjournalVolNumPages{IEEE Security \& Privacy}{15}{3}{90--95}.
\newblock
\begin{APACrefURL} {\url{https://doi.org/10.1109/MSP.2017.54}} \end{APACrefURL}
\newblock

\newblock

\PrintBackRefs{\CurrentBib}

\bibitem [\protect \citeauthoryear {%
Yamin%
, Katt%
\BCBL {}\ \BBA {} Gkioulos%
}{%
Yamin%
\ \protect \BOthers {.}}{%
{\protect \APACyear {2020}}%
}]{%
yamin2020}
\APACinsertmetastar {%
yamin2020}%
\begin{APACrefauthors}%
Yamin, M.M.%
, Katt, B.%
\BCBL {} Gkioulos, V.%
\end{APACrefauthors}%
\unskip\
\newblock
\APACrefYearMonthDay{2020}{}{}.
\newblock
{\BBOQ}\APACrefatitle {Cyber ranges and security testbeds: Scenarios,
  functions, tools and architecture} {Cyber ranges and security testbeds:
  Scenarios, functions, tools and architecture}.{\BBCQ}
\newblock
\APACjournalVolNumPages{Computers \& Security}{88}{}{101636}.
\newblock
\begin{APACrefURL} {\url{https://doi.org/10.1016/j.cose.2019.101636}}
  \end{APACrefURL}
\newblock

\newblock

\PrintBackRefs{\CurrentBib}

\bibitem [\protect \citeauthoryear {%
Yett%
\ \protect \BOthers {.}}{%
Yett%
\ \protect \BOthers {.}}{%
{\protect \APACyear {2020}}%
}]{%
Yett2020}
\APACinsertmetastar {%
Yett2020}%
\begin{APACrefauthors}%
Yett, B.%
, Snyder, C.%
, Zhang, N.%
, Hutchins, N.%
, Mishra, S.%
\BCBL {} Biswas, G.%
\end{APACrefauthors}%
\unskip\
\newblock
\APACrefYearMonthDay{2020}{}{}.
\newblock
{\BBOQ}\APACrefatitle {Using Log and Discourse Analysis to Improve
  Understanding of Collaborative Programming} {Using log and discourse analysis
  to improve understanding of collaborative programming}.{\BBCQ}
\newblock
 \APACrefbtitle {{Proceedings of the 28th International Conference on Computers
  in Education}} {{Proceedings of the 28th International Conference on
  Computers in Education}}\ (\BPGS\ 1--10).
\newblock
\begin{APACrefURL}
  {\url{https://apsce.net/icce/icce2020/proceedings/paper_158.pdf}}
  \end{APACrefURL}
\PrintBackRefs{\CurrentBib}

\bibitem [\protect \citeauthoryear {%
Zeng%
, Deng%
, Hsiao%
, Huang%
\BCBL {}\ \BBA {} Chung%
}{%
Zeng%
\ \protect \BOthers {.}}{%
{\protect \APACyear {2018}}%
}]{%
Zeng2018improving}
\APACinsertmetastar {%
Zeng2018improving}%
\begin{APACrefauthors}%
Zeng, Z.%
, Deng, Y.%
, Hsiao, I.%
, Huang, D.%
\BCBL {} Chung, C\BHBI J.%
\end{APACrefauthors}%
\unskip\
\newblock
\APACrefYearMonthDay{2018}{10}{}.
\newblock
{\BBOQ}\APACrefatitle {{Improving student learning performance in a virtual
  hands-on lab system in cybersecurity education}} {{Improving student learning
  performance in a virtual hands-on lab system in cybersecurity
  education}}.{\BBCQ}
\newblock
 \APACrefbtitle {{2018 IEEE Frontiers in Education Conference (FIE)}} {{2018
  IEEE Frontiers in Education Conference (FIE)}}\ (\BPG~1–5).
\newblock
\APACaddressPublisher{New York, NY, USA}{IEEE}.
\newblock
\begin{APACrefURL} {\url{https://doi.org/10.1109/FIE.2018.8658855}}
  \end{APACrefURL}
\PrintBackRefs{\CurrentBib}

\bibitem [\protect \citeauthoryear {%
Zurita%
\ \protect \BOthers {.}}{%
Zurita%
\ \protect \BOthers {.}}{%
{\protect \APACyear {2020}}%
}]{%
Zurita2020}
\APACinsertmetastar {%
Zurita2020}%
\begin{APACrefauthors}%
Zurita, G.%
, Shukla, A.K.%
, Pino, J.A.%
, Merigó, J.M.%
, Lobos-Ossandón, V.%
\BCBL {} Muhuri, P.K.%
\end{APACrefauthors}%
\unskip\
\newblock
\APACrefYearMonthDay{2020}{}{}.
\newblock
{\BBOQ}\APACrefatitle {A bibliometric overview of the Journal of Network and
  Computer Applications between 1997 and 2019} {A bibliometric overview of the
  journal of network and computer applications between 1997 and 2019}.{\BBCQ}
\newblock
\APACjournalVolNumPages{Journal of Network and Computer
  Applications}{165}{}{102695}.
\newblock
\begin{APACrefURL} {\url{https://doi.org/10.1016/j.jnca.2020.102695}}
  \end{APACrefURL}
\newblock

\newblock

\PrintBackRefs{\CurrentBib}

\end{thebibliography}

\end{document}